# Fundamentals of optical non-reciprocity based on optomechanical coupling

Mohammad-Ali Miri[1], Freek Ruesink[2], Ewold Verhagen[2], and Andrea Alù[1,*]

[1]Department of Electrical and Computer Engineering, The University of Texas at Austin, Austin, Texas 78712, USA

[2]Center for Nanophotonics, FOM Institute AMOLF, Science Park 104, 1098 XG Amsterdam, The Netherlands

[*]Email: alu@mail.utexas.edu

**Abstract**

*Optical isolation, non-reciprocal phase transmission and topological phases for light based on synthetic gauge fields have been raising significant interest in the recent literature. Cavity-optomechanical systems that involve two optical modes coupled to a common mechanical mode form an ideal platform to realize these effects, providing the basis for various recent demonstrations of optomechanically induced non-reciprocal light transmission. Here, we establish a unifying theoretical framework to analyze optical non-reciprocity and breaking of time-reversal symmetry in multimode optomechanical systems. We highlight two general scenarios to achieve isolation, relying on either optical or mechanical losses. Depending on the loss mechanism, our theory defines the ultimate requirements for optimal isolation and the available operational bandwidth in these systems. We also analyze the effect of sideband resolution on the performance of optomechanical isolators, highlighting the fact that non-reciprocity can be preserved even in the unresolved sideband regime. Our results provide general insights into a broad class of parametrically modulated non-reciprocal devices, paving the way towards optimal non-reciprocal systems for low-noise integrated nanophotonics.*

## 1. Introduction

Non-reciprocal elements are crucial in nanophotonic communication systems. Such devices allow the transmission of signals in one direction while blocking those propagating in the opposite one, avoiding interference and protecting optical sources. In general, achieving non-reciprocity requires breaking the time-reversal symmetry inherent in the governing electromagnetic wave equations, a symmetry that holds as long as the structure is linear, time invariant, and it is not biased by a quantity that is odd under time reversal. In practice, optical isolation is commonly achieved based on the magneto-optic effect [1], i.e., by applying a static magnetic bias. However, such devices tend to be bulky, costly and not CMOS-compatible, motivating the on-going search for alternative strategies to break reciprocity in chip-scale devices. Over the last few years, several approaches have been suggested in integrated photonic systems. Examples include nonlinear structures with a spatially asymmetric refractive index profile [2] and systems that undergo a dynamic spatio-temporal modulation of the refractive index profile, thus mimicking the effect of an external gauge bias and inducing non-reciprocal behavior [3]-[6]. Microring resonators with a traveling wave index modulation, acting as an angular momentum bias, have been proposed as an efficient way to break reciprocity in compact devices [6]-[7], a concept that has been realized in a discretized arrangement of resonators with out-of-phase temporal modulations [8]-[9]. Recently, it has been realized that optomechanical coupling can also be used to impart the required form of synthetic gauge required to induce optical non-reciprocity [10]-[19]. In this context, different theories have been presented to describe possible optomechanical implementations of on-chip isolators [11]-[12],[17]-[18].

Here we present a general theoretical framework to describe multimode optomechanical arrangements for non-reciprocal transmission, establishing a minimal model that captures the essential mechanisms behind the operation of the different geometries discussed in the recent literature [11]-[12],[17]-[18]. We show that optomechanically-induced non-reciprocity can be observed in a wide class of multimode systems, as long as a minimum set of necessary and sufficient conditions are satisfied. These conditions are expressed in terms of the mode-port coupling matrix of the underlying optical system, as well as the relative phases and intensities of the driving lasers used to bias.

Previously reported geometries [11]-[12],[18] are then discussed as specific cases of our general theory. We derive fundamental conditions to achieve non-reciprocal responses in phase and intensity, and discuss the requirements to maximize isolation, non-reciprocal phase difference and bandwidth constraints. We also investigate the performance of optomechanical isolators and gyrators in both resolved and unresolved sideband regimes, and under linear and nonlinear conditions.

The paper is organized as follows. In Section 2, we review the temporal coupled mode theory of a general two-port optical system that involves two modes and derive the minimal requirements for non-reciprocity, showing the general necessity of non-reciprocal mode conversion. Next, we show how a mechanical mode coupled to both optical modes can mediate such non-reciprocal conversion, and we derive the conditions for the optical drive fields to optimally break reciprocity. Section 4 explains how such conversion can lead to non-reciprocal phase shifting and isolation in two classes of implementations, based on end- and side-coupled resonator geometries respectively, which differ in the loss mechanism responsible for isolation. Sections 5 and 6 study how transmission through both classes of systems depends on the geometry and the drive fields. In both cases, the conditions for ideal isolation are derived, and their realization in terms of the involved parameters is discussed. In Section 7, we explore the possibility of non-reciprocal amplification. Section 8 is then devoted to the extension of this treatment to a more general scenario in which both sidebands are taken into account, pointing out the relevant fact that sideband resolution is not necessary to yield non-reciprocal transmission. The linear eigenmodes of the system are explored in Section 9, allowing a rigorous study of the instability threshold for these devices. The steady-state biasing conditions are then investigated in Section 10, followed by rigorous time-domain simulations of the governing nonlinear dynamical equations that validate our results in specific sample geometries.

## 2. Coupled mode theory and time-reversal symmetry breaking in a two-port/two-mode optical system

Before investigating the hybrid optomechanical system at the core of this paper, consider a general optical two-port/two-mode system as shown in Fig. 1, which can be described through the coupled mode formalism [20]

$$\frac{d}{dt}\begin{pmatrix}a_1\\a_2\end{pmatrix} = i\mathcal{M}\begin{pmatrix}a_1\\a_2\end{pmatrix} + B\begin{pmatrix}s_1^+\\s_2^+\end{pmatrix}, \quad (1)$$

$$\begin{pmatrix}s_1^-\\s_2^-\end{pmatrix} = C\begin{pmatrix}s_1^+\\s_2^+\end{pmatrix} + D\begin{pmatrix}a_1\\a_2\end{pmatrix}, \quad (2)$$

where $a_{1,2}$ are the amplitudes of the two modes and $s_{1,2}^\pm$ represent the incoming (+) and outgoing (−) signals at the two ports. The matrix $C$ describes the direct path scattering matrix between the two ports, while $D$ and $B$ describe the port to mode and mode to port coupling processes, respectively. Finally, $\mathcal{M}$ represents a linear evolution matrix of the optical modes in the absence of excitation. Here we assume that the evolution operator does not depend explicitly on time, as in the case of systems with externally controlled parametric modulation. However, $\mathcal{M}$ can include time derivatives, which is the case for an optomechanical system involving self-induced parametric modulation. In such systems, $\mathcal{M}$ can be decomposed in two terms, one describing the bare optical system, $\Theta$, and a second term associated with optomechanical interactions. In general, the bare optical evolution operator can be written as $\Theta = O + \frac{i}{2}K$ where $O$ and $K$, both being real and symmetric matrices, represent resonance and damping frequencies. The diagonal and off-diagonal elements of $O$ represent respectively the resonance frequencies of the two optical modes $(\omega_1, \omega_2)$ and the mutual coupling between the two modes $(\mu)$. The losses, on the other hand, can be decomposed into exchange $(K_e)$ and intrinsic losses $(K_\ell)$ as $K = K_e + K_\ell$ (in a conservative treatment of the system, one can consider the intrinsic losses as extra ports that work as leakage channels). The diagonal and off-diagonal elements of $K$ respectively represent the total losses of each modes $(\kappa_1, \kappa_2)$ and the coupling between two modes due to interference in the joint output channels $(\kappa_r)$. Without loss of generality, here we assume an eigenbasis that diagonalizes the bare optical evolution matrix $\Theta$. In doing so, the diagonal elements of $\Theta$ can be written as $\omega_{1,2} + i\kappa_{1,2}/2$, where $\omega_{1,2} = \omega_0 \mp \mu$ represent the resonance frequencies of the two modes, $\kappa_{1,2}$ their total losses and $2\mu$ a possible frequency splitting. In addition, we define leakage coefficients $\eta_{1,2}$, which describe the ratios of external losses (due to decay into the considered ports) to total losses of each mode, i.e., $\kappa_{e_{1,2}} = \eta_{1,2}\kappa_{1,2}$ and $\kappa_{\ell_{1,2}} = (1 - \eta_{1,2})\kappa_{1,2}$.

The matrices involved in (1,2) are not independent, as time-reversal symmetry and energy conservation impose relevant restrictions on them. We use the convention in which each optical mode is explicitly coupled to the input/output channels in a reciprocal fashion, meaning $B = D^T$. Then, $\det(CD^* + D) = 0$, and $D^\dagger D = K_e$, where in these relations "T" and "†" respectively represent the transpose and conjugated transpose operations [20]. Based on these relations, we can derive a general condition on the determinant of the coupling matrix $D$: since $D^\dagger D = K_e$, we can write $|\det(D)|^2 = \det(K_e) = \eta_1\eta_2\kappa_1\kappa_2$. Using $CD^* = -D$, we find that $\det(C)\det(D)^* = \det(D)$. Here $C$ is a unitary matrix, thus $|\det(C)| = 1$. In general, nothing can be said about the phase of $\det(C)$. However, by properly choosing the reference plane at one of the ports, we can control this phase and, without loss of generality, we assume in the following that $\det(C) = -1$, yielding $\det(D) = i\sqrt{\eta_1\eta_2\kappa_1\kappa_2}$.

In the frequency domain, the scattering matrix of a system governed by Eqs. (1,2), defined as

$$\begin{pmatrix} s_1^- \\ s_2^- \end{pmatrix} = S(\omega) \begin{pmatrix} s_1^+ \\ s_2^+ \end{pmatrix}, \quad (3)$$

can be written as

$$S = C + iD(M(\omega) + \omega I)^{-1}D^T. \quad (4)$$

Based on this relation the difference between forward and backward transmission, which quantifies non-reciprocity, can be written in a very compact and general form:

$$S_{21} - S_{12} = i\frac{\det(D)(m_{12}-m_{21})}{\det(M(\omega)+\omega I)}, \quad (5)$$

which is a fundamental relation for the rest of this work. According to this expression, two conditions are necessary and sufficient to break reciprocity in a general two-port optical system based on two coupled optical modes [17]: (a) $\det(D) \neq 0$, and (b) $m_{12} \neq m_{21}$. The full rank of the coupling matrix $D$ can be ensured with a suitable asymmetry in the coupling of the two modes to the two ports, i.e., $d_{11}/d_{21} \neq d_{12}/d_{22}$. The second condition, on the other hand, is quite demanding, as in a linear, time-invariant, time-reversible system the evolution matrix is always symmetric. In the next section, we show that optomechanical interactions, when properly controlled, can break the symmetry of the effective evolution matrix, thus enabling optical non-reciprocity.

## 3. Multimode cavity optomechanical system

### 3.1. Optomechanical evolution equations

Consider the case in which the general system discussed in the previous section supports a single mechanical mode coupled to both optical modes. The effective mass, resonance frequency and decay rate of the mechanical mode are $m$, $\Omega_m$ and $\Gamma_m$, respectively, while the optical modes' frequency shift per mechanical displacement are $\mathcal{G}_1$ and $\mathcal{G}_2$, respectively. In the frame of control frequency $\omega_L$ the evolution of this system is described by

$$\frac{d}{dt}\begin{pmatrix}a_1\\a_2\end{pmatrix} = i\begin{pmatrix}\Delta_1 + \mathcal{G}_1 x + i\kappa_1/2 & 0 \\ 0 & \Delta_2 + \mathcal{G}_2 x + i\kappa_2/2\end{pmatrix}\begin{pmatrix}a_1\\a_2\end{pmatrix} + D^T\begin{pmatrix}s_1^+\\s_2^+\end{pmatrix}, \quad (6)$$

$$\frac{d^2}{dt^2}x = -\Omega_m^2 x - \Gamma_m \frac{d}{dt}x + \frac{\hbar}{m}(\mathcal{G}_1|a_1|^2 + \mathcal{G}_2|a_2|^2), \quad (7)$$

where $x$ is the position of the mechanical resonator with respect to its reference point. Here, $\Delta_{1,2} = \omega_L - \omega_{1,2}$ represent the detuning of the resonance frequencies with respect to the driving frequency. Assuming $\omega_1 = \omega_0 - \mu$ and $\omega_2 = \omega_0 + \mu$, we can write $\Delta_1 = \Delta + \mu$ and $\Delta_2 = \Delta - \mu$, where $\Delta = \omega_L - \omega_0$ is a detuning from the center of the two resonance frequencies.

### 3.2. Linearized optomechanical system and scattering parameters

Assuming that the optical modes are strongly driven by a control signal at $\omega_L$, the evolution equations can be linearized for weak probes at $\omega_p = \omega_L + \omega$. In this case, the modal optical amplitudes and the mechanical displacements can be written as $a_{1,2}(t) = \bar{a}_{1,2} + \delta a_{1,2}(t)$ and $x(t) = \bar{x} + \delta x(t)$ where $|\delta a_{1,2}| \ll |\bar{a}_{1,2}|$. Here $\bar{a}_{1,2}$ and $\bar{x}$ are the fixed point biases of the optical and mechanical resonators, which are obtained from Eqs. (6,7) at steady state, i.e., for $d/dt \to 0$. The evolution of the modulating optical $\delta a_{1,2}$ and mechanical $\delta x$ signals is governed by the linearized equations

$$\frac{d}{dt}\begin{pmatrix}\delta a_1\\ \delta a_2\end{pmatrix} = i\begin{pmatrix}\bar{\Delta}_1 + i\kappa_1/2 & 0\\ 0 & \bar{\Delta}_2 + i\kappa_2/2\end{pmatrix}\begin{pmatrix}\delta a_1\\ \delta a_2\end{pmatrix} + i\begin{pmatrix}G_1\\ G_2\end{pmatrix}\delta x + D^T\begin{pmatrix}\delta s_1^+\\ \delta s_2^+\end{pmatrix}, \quad (8)$$

$$\frac{d^2}{dt^2}\delta x = -\Omega_m^2 \delta x - \Gamma_m \frac{d}{dt}\delta x + \frac{\hbar}{m}(G_1^* \delta a_1 + G_1 \delta a_1^* + G_2^* \delta a_2 + G_2 \delta a_2^*), \quad (9)$$

where $\bar{\Delta}_{1,2} = \Delta_{1,2} + \mathcal{G}_{1,2}\bar{x}$ are the modified frequency detuning factors, and $G_{1,2} = \mathcal{G}_{1,2}\bar{a}_{1,2}$ are the enhanced optomechanical frequency shifts. Here, we assume both modes being driven in the extreme red- or blue-detuned regimes, i.e., $\bar{\Delta}_{1,2} \approx \mp\Omega_m$. In addition, in this section we assume for now a sideband resolved operation, i.e., the mechanical frequency is larger than the optical linewidths, $\Omega_m > \kappa_{1,2}$. Under these conditions, and for a probe signal approximately centered at the optical resonance frequency, it is possible to show that the terms with complex conjugate fields in the above equations can be ignored [21]. We will lift the sideband resolution assumption in section 8.

The dynamical equations (8,9) for the optical modes can be written in the form of Eq. (1) as

$$\frac{d}{dt}\begin{pmatrix}\delta a_1\\ \delta a_2\end{pmatrix} = i\left[\begin{pmatrix}\bar{\Delta}_1 + i\frac{\kappa_1}{2} & 0\\ 0 & \bar{\Delta}_2 + i\frac{\kappa_2}{2}\end{pmatrix} + \frac{\hbar}{m\left(\frac{d^2}{dt^2}+\Gamma_m\frac{d}{dt}+\Omega_m^2\right)}\begin{pmatrix}|G_1|^2 & G_1 G_2^*\\ G_1^* G_2 & |G_2|^2\end{pmatrix}\right]\begin{pmatrix}\delta a_1\\ \delta a_2\end{pmatrix} + D^T\begin{pmatrix}\delta s_1^+\\ \delta s_2^+\end{pmatrix},$$

(10)

Therefore, in the frequency domain (here, the Fourier transform is defined as $a_{1,2}(\omega) = \int a_{1,2}(t)e^{i\omega t}d\omega$, where again $\omega = \omega_p - \omega_L$ represents the probe frequency evaluated with respect to the control frequency), we have:

$$i\left[\begin{pmatrix}\omega + \bar{\Delta}_1 + i\frac{\kappa_1}{2} & 0\\ 0 & \omega + \bar{\Delta}_2 + i\frac{\kappa_2}{2}\end{pmatrix} - \frac{\hbar}{m(\omega^2-\Omega_m^2+i\Gamma_m\omega)}\begin{pmatrix}|G_1|^2 & G_1 G_2^*\\ G_1^* G_2 & |G_2|^2\end{pmatrix}\right]\begin{pmatrix}\delta a_1\\ \delta a_2\end{pmatrix} + D^T\begin{pmatrix}\delta s_1^+\\ \delta s_2^+\end{pmatrix} = 0,$$

(11)

and the evolution operator can thus be written as

$$M = \begin{pmatrix} \bar{\Delta}_1 + i\kappa_1/2 & 0 \\ 0 & \bar{\Delta}_2 + i\kappa_2/2 \end{pmatrix} - \frac{\hbar}{\Sigma_m} \begin{pmatrix} |G_1|^2 & G_1 G_2^* \\ G_1^* G_2 & |G_2|^2 \end{pmatrix}, \quad (12)$$

where $\Sigma_m = m(\omega^2 - \Omega_m^2 + i\Gamma_m\omega)$ represents the inverse mechanical susceptibility. As this relation clearly shows, the symmetry of the evolution matrix can be broken through the optomechanical interaction terms, as long as $G_1 G_2^* \neq G_1^* G_2$ (see Fig. 2). Assuming a phase difference $\Delta\phi = \sphericalangle G_2 - \sphericalangle G_1$ between the enhanced optomechanical frequency shifts, this latter condition requires $\Delta\phi \neq n\pi$ where $n = 0, \pm 1, \pm 2, \ldots$ A similar conclusion can be reached analyzing directly the scattering matrix (4), which leads to

$$S = C + iD \begin{pmatrix} \Sigma_{o_1} - \frac{\hbar}{\Sigma_m}|G_1|^2 & -\frac{\hbar}{\Sigma_m} G_1 G_2^* \\ -\frac{\hbar}{\Sigma_m} G_1^* G_2 & \Sigma_{o_2} - \frac{\hbar}{\Sigma_m}|G_2|^2 \end{pmatrix}^{-1} D^T, \quad (13)$$

where $\Sigma_{o_{1,2}} = \left(\omega + \bar{\Delta}_{1,2} + i\kappa_{1,2}/2\right)$ represents the inverse optical susceptibility of the two optical modes. The scattering coefficients can be then explicitly obtained:

$$S_{11} = c_{11} + i \frac{d_{12}^2(\Sigma_{o_1}\Sigma_m - \hbar|G_1|^2) + d_{11}^2(\Sigma_{o_2}\Sigma_m - \hbar|G_2|^2) + d_{11}d_{12}\hbar(G_1 G_2^* + G_1^* G_2)}{\Sigma_{o_1}\Sigma_{o_2}\Sigma_m - \hbar(\Sigma_{o_2}|G_1|^2 + \Sigma_{o_1}|G_2|^2)}, \quad (14.a)$$

$$S_{12} = c_{12} + i \frac{d_{12}d_{22}(\Sigma_{o_1}\Sigma_m - \hbar|G_1|^2) + d_{11}d_{21}(\Sigma_{o_2}\Sigma_m - \hbar|G_2|^2) + d_{11}d_{22}\hbar G_1 G_2^* + d_{12}d_{21}\hbar G_1^* G_2}{\Sigma_{o_1}\Sigma_{o_2}\Sigma_m - \hbar(\Sigma_{o_2}|G_1|^2 + \Sigma_{o_1}|G_2|^2)}, \quad (14.b)$$

$$S_{21} = c_{21} + i \frac{d_{12}d_{22}(\Sigma_{o_1}\Sigma_m - \hbar|G_1|^2) + d_{11}d_{21}(\Sigma_{o_2}\Sigma_m - \hbar|G_2|^2) + d_{12}d_{21}\hbar G_1 G_2^* + d_{11}d_{22}\hbar G_1^* G_2}{\Sigma_{o_1}\Sigma_{o_2}\Sigma_m - \hbar(\Sigma_{o_2}|G_1|^2 + \Sigma_{o_1}|G_2|^2)}, \quad (14.c)$$

$$S_{22} = c_{22} + i \frac{d_{22}^2(\Sigma_{o_1}\Sigma_m - \hbar|G_1|^2) + d_{21}^2(\Sigma_{o_2}\Sigma_m - \hbar|G_2|^2) + d_{21}d_{22}\hbar(G_1 G_2^* + G_1^* G_2)}{\Sigma_{o_1}\Sigma_{o_2}\Sigma_m - \hbar(\Sigma_{o_2}|G_1|^2 + \Sigma_{o_1}|G_2|^2)}. \quad (14.d)$$

Using Eq. (5) and the determinant relation, the complex difference between forward and backward transmission coefficients becomes

$$S_{21} - S_{12} = -2i\sqrt{\eta_1\eta_2\kappa_1\kappa_2} \frac{\hbar|G_1||G_2|\sin(\Delta\phi)}{\Sigma_{o_1}\Sigma_{o_2}\Sigma_m - \hbar(\Sigma_{o_2}|G_1|^2 + \Sigma_{o_1}|G_2|^2)}. \quad (15)$$

This general relation ensures that the maximum contrast between forward and backward transmission coefficients is obtained when the driving fields are in quadrature, $\Delta\phi = \pm\pi/2$.

## 4. Optomechanically induced non-reciprocity

### 4.1. Fabry-Peròt model

In order to provide an intuitive understanding of the underlying physics involved in the design of a non-reciprocal optomechanical system, we consider two Fabry-Peròt models. These are referred to as end- and side-coupled structures (Figs. 3(a,c) and (b,d), respectively), in analogy with their integrated photonic counterparts that will be introduced later. The difference between these systems is a direct light propagation path between the two input and output ports in scenarios (b) and (d), which is absent in (a) and (c). For both systems (Eq. (12)), the mechanically mediated hopping rate from mode 1 to 2 reads $\mu_m^{1 \to 2} = -\hbar G_1 G_2^* / \Sigma_m(\omega)$, while for the opposite process $\mu_m^{2 \to 1} = -\hbar G_1^* G_2 / \Sigma_m(\omega)$. At resonance, and for $\Delta\phi = \pi/2$, this coupling reduces to $\mu_m^{1 \to 2} = \hbar |G_1||G_2|/m\Gamma_m\Omega_m$ and $\mu_m^{2 \to 1} = -\mu_m^{1 \to 2}$, which reveals that this coupling imprints opposite phase for oppositely traveling photons. However, in order to obtain isolation this non-reciprocal mode transfer path needs to be interfered with a second optical path.

In the end-coupled structure, such an additional path is provided by direct hopping between the optical modes at rate $\mu$. A finite optical coupling ($\mu \neq 0$) allows one-way destructive interference between the two paths, resulting in isolation. Critically, in order to create complete destructive interference between the two paths, a careful match between hopping rates is required. Optimal isolation in the end-coupled geometry therefore occurs for $\mu = |\mu_m|$, which is consistent with the condition derived in [18] following a different theoretical approach. At first sight, this result seems to suggest that it is possible to equally increase or decrease both $\mu$ and $|\mu_m|$ to achieve ideal isolation. However, careful inspection of the underlying equations, as detailed in section 5, will show that there is an optimum value for $\mu$, related to the rate at which photons are lost through the mechanical loss channel.

In contrast, the side-coupled geometry (Fig. 3(b)) can be seen as the end-coupled system of Fig. 3(a) positioned in an optical interferometer. In this case, a direct propagation channel provides the path with which the mode-transfer processes can interfere, external to the cavities. Considering the direct channel to be lossless, one can intuitively understand that complete destructive interference happens when all the light entering the optomechanical system at cavity 1 exits at cavity 2. In other words, complete isolation is achieved for ideal mode-transfer, which occurs for $|G_1||G_2| \to \infty$.

Although the Fabry-Peròt models introduced here provide an intuitive understanding of the major processes leading to non-reciprocal light transmission in the general platform analysed in this paper, a more quantitative discussion based on Eqs. (14,15) requires the implementation of system specific $D$-matrices, which are derived in the next section.

### 4.2. Integrated photonic geometries

The Fabry-Perot models introduced in the previous sub-section can be modeled in abstract waveguide representations as in Fig. 4. We assume that the optical cavity depicted here supports two modes, which is equivalent to considering a coupled pair of single mode cavities, and that the optical cavity exhibits mirror symmetry with respect to a plane orthogonal to the propagation direction, thus supporting modes with even and odd symmetry. In a unified treatment of both the end- and side-coupled scenarios, we write the coupled mode equations in the eigenbasis of the normal modes, which diagonalizes the bare optical evolution matrix.

In the end-coupled geometry (Fig. 4(a)), we assume that the only propagation path between the two ports is through the optical resonators, such that the direct scattering matrix $C$ reads

$$C = \begin{pmatrix} i & 0 \\ 0 & i \end{pmatrix}, \quad (16)$$

where the arbitrary phase of the reflection coefficient is chosen to ensure $\det(C) = -1$. The symmetry of the modes dictates $d_{11} = d_{21}$ (even) and $d_{12} = -d_{22}$ (odd) (see the inset of Fig. 4). Using these considerations and given that $D^\dagger D = K_e$, we obtain $|d_{11}|^2 = \eta_1 \kappa_1/2$ and $|d_{22}|^2 = \eta_2 \kappa_2/2$. Together with the condition $CD^* = -D$, the coupling matrix is thus fully determined as

$$D = \frac{e^{-i\pi/4}}{\sqrt{2}} \begin{pmatrix} \sqrt{\eta_1 \kappa_1} & -\sqrt{\eta_2 \kappa_2} \\ \sqrt{\eta_1 \kappa_1} & \sqrt{\eta_2 \kappa_2} \end{pmatrix}. \quad (17)$$

In contrast, when the optical cavity supporting two modes is side-coupled to a bus waveguide (Fig. 4(b)), the direct path scattering matrix without any reflection reads

$$C = \begin{pmatrix} 0 & 1 \\ 1 & 0 \end{pmatrix}, \quad (18)$$

ensuring the same condition $\det(C) = -1$. Using a similar procedure, the coupling matrix for the side-coupled geometry is obtained as

$$D = \frac{1}{\sqrt{2}} \begin{pmatrix} i\sqrt{\eta_1 \kappa_1} & -\sqrt{\eta_2 \kappa_2} \\ i\sqrt{\eta_1 \kappa_1} & \sqrt{\eta_2 \kappa_2} \end{pmatrix}. \quad (19)$$

In the next sections, we apply this analytical model to analyze the general conditions for non-reciprocity in these two integrated photonic schemes.

## 5- End-coupled structure

The scattering parameters of the end-coupled geometry are provided in Eqs. (14,16,17). For two optical modes that exhibit the same amount of intrinsic and external losses ($\eta_1 = \eta_2 \equiv \eta$, $\kappa_1 = \kappa_2 \equiv \kappa$), and are equally driven ($|G_1| = |G_2| \equiv |G|$), Eqs. (14) reduce to:

$$S_{11} = i + \eta\kappa \frac{\Sigma_o \Sigma_m - \hbar|G|^2(1+\cos(\Delta\phi))}{(\Sigma_o^2 - \mu^2)\Sigma_m - 2\hbar|G|^2 \Sigma_o}, \quad (20.a)$$

$$S_{12} = \eta\kappa \frac{\mu \Sigma_m - i\hbar|G|^2 \sin(\Delta\phi)}{(\Sigma_o^2 - \mu^2)\Sigma_m - 2\hbar|G|^2 \Sigma_o}, \quad (20.b)$$

$$S_{21} = \eta\kappa \frac{\mu \Sigma_m + i\hbar|G|^2 \sin(\Delta\phi)}{(\Sigma_o^2 - \mu^2)\Sigma_m - 2\hbar|G|^2 \Sigma_o}, \quad (20.c)$$

$$S_{22} = i + \eta\kappa \frac{\Sigma_o \Sigma_m - \hbar|G|^2(1-\cos(\Delta\phi))}{(\Sigma_o^2 - \mu^2)\Sigma_m - 2\hbar|G|^2 \Sigma_o}. \quad (20.d)$$

where we have used $\Sigma_{o_{1,2}} = \Sigma_o \pm \mu$, where $\Sigma_o = \omega + \bar{\Delta} + i\kappa/2$ and $2\mu$ represents the resonance frequency splitting of the two optical modes. These relations again show that the contrast between $S_{12}$ and $S_{21}$ is maximal for $\Delta\phi = \pi/2$. Interestingly, under this pump condition the reflection coefficients $S_{11}$ and $S_{22}$ are equal, i.e., the transmission difference is not induced by asymmetric mismatch at the port, but by asymmetric absorption. On the other hand, for $\Delta\phi = 0$ reciprocity is restored ($S_{12} = S_{21}$), while the reflection coefficients are no longer equal. Any other phase difference provides asymmetry in both transmission and reflection, and non-optimal isolation.

Figure 5 shows the scattering parameters of an end-coupled structure, when driven in the extreme red-detuning regime $\bar{\Delta} = -\Omega_m$ for different incident control amplitudes and changing drive phase $\Delta\phi$. As expected, an in-phase drive ($\Delta\phi = 0$) results in a reciprocal system, while asymmetric driving ($\Delta\phi = \pi/2$) results in non-reciprocal transmission around the optical resonance $\omega = \Omega_m$. Interestingly, the contrast between forward and backward transmission approaches zero at both low and high power driving regimes, consistent with the fact that maximum contrast is expected for $\mu = |\mu_m|$. The relatively low values of transmissivities depicted in this figure are due to the fact that we assume equal intrinsic and external losses ($\eta = 1/2$). In principle, the transmissivities can be increased up to unity for $\eta \to 1$. In these plots, we chose $\eta = 1/2$ to enable a direct comparison with the side-coupled geometry in the next section.

### 5.1. Degenerate modes: optical gyrator

An interesting scenario arises when the two optical modes are degenerate ($\mu = 0$). This implies absence of direct coupling between them, such that the only coupling path between the two ports is through the mechanical mode. In this scenario, the transmission coefficients are simplified to

$$S_{12} = -\eta\kappa \frac{i\hbar|G|^2 \sin(\Delta\phi)}{\Sigma_o^2 \Sigma_m - 2\hbar|G|^2 \Sigma_o}, \quad (21.a)$$

$$S_{21} = +\eta\kappa \frac{i\hbar|G|^2 \sin(\Delta\phi)}{\Sigma_o^2 \Sigma_m - 2\hbar|G|^2 \Sigma_o}. \quad (21.b)$$

According to this relation, the amplitudes of the forward and backward transmission coefficients are equal, but exhibit opposite phase. This structure thus operates as a gyrator, i.e., a non-reciprocal phase shifter with phase difference equal to $\pi$. The intensity and phase of the transmission coefficients of this system are shown in Fig. 6, highlighting an increase in transmission bandwidth when the pump power increases. Interestingly, the difference between phases of the forward and backward transmission coefficients is independent of frequency, even though the amplitude response is governed by the optomechanical lineshape.

The phase difference of $\pi$ between forward and backward probes arises under the assumption that even and odd modes are pumped with equal intensity. In principle, however, the phase difference can be controlled through an unbalanced pumping. In this case, by assuming equal losses for the modes, it is straightforward to show

$$\frac{S_{12}}{S_{21}} = \frac{|G_1|^2-|G_2|^2-i2|G_1||G_2|\sin(\Delta\phi)}{|G_1|^2-|G_2|^2+i2|G_1||G_2|\sin(\Delta\phi)}, \quad (22)$$

which clearly shows the controllability of the non-reciprocal phase via the enhanced optomechnaical coupling coefficients $G_{1,2} = \mathcal{G}_{1,2}\bar{a}_{1,2}$. The relation between port excitations $\bar{s}_{1,2}$ and mode biases $\bar{a}_{1,2}$ is further discussed in section 10.

### *5.2. Conditions for ideal isolation*

In this sub-section we are ready to explore the conditions for optimal isolation in this end-coupled geometry, i.e., $S_{12} = 0$ and $|S_{21}| = 1$. Assuming $\Delta\phi = \pi/2$, $\bar{\Delta} = -\Omega_m$ and $\omega = \Omega_m$, the transmission coefficients in Eqs. (20) reduce to

$$S_{12}(\omega = \Omega_m) = -2\eta \frac{\frac{\mu}{\kappa/2}-\mathcal{C}}{1+\left(\frac{\mu}{\kappa/2}\right)^2+2\mathcal{C}}, \quad (23.\text{a})$$

$$S_{21}(\omega = \Omega_m) = -2\eta \frac{\frac{\mu}{\kappa/2}+\mathcal{C}}{1+\left(\frac{\mu}{\kappa/2}\right)^2+2\mathcal{C}}, \quad (23.\text{b})$$

where

$$\mathcal{C}_1 = \mathcal{C}_2 = \mathcal{C} = \frac{\hbar|G|^2}{2m\Omega_m(\Gamma_m/2)(\kappa/2)} \quad (24)$$

represents the multi-photon cooperativity of each optical mode. According to these relations, and consistent with the discussion in the previous section, complete rejection of the backward propagating probe requires a balance between the normalized mode splitting and total cooperativity:

$$\frac{2\mu}{\kappa} = \mathcal{C}. \quad (25)$$

This can be understood from the fact that the direct optical mode coupling, occurring at an energy transfer rate $\mu$, should completely cancel the mechanically-mediated conversion at rate $C\kappa/2 = \hbar|G|^2/m\Omega_m\Gamma_m$. Under this condition, the forward transmission becomes

$$|S_{21}(\omega = \Omega_m)| = \frac{4\eta C}{(C+1)^2}, \quad (26)$$

which is generally less than unity, implying a non-zero insertion loss. Asymptotically low ($C \ll 1$) and high ($C \gg 1$) values of cooperativity yield zero forward transmission, and maximum transmission is obtained for $C = 1$, which results in $\max(|S_{21}|) = \eta$. As expected, complete forward transmission and zero insertion loss can be achieved when the optical modes have zero absorption, i.e., $\kappa_\ell = 0$, or equivalently, $\eta = 1$. According to Eq. (25), in order to simultaneously block the backward probe, one needs to enforce $2\mu = \kappa$. Figure 7(a) shows the transmission contrast in a contour map versus the normalized mode splitting (horizontal axis) and cooperativity (vertical axis) for $\eta = 1$.

Although the above analysis implies that it is feasible to achieve ideal isolation in a system with no optical absorption, isolation in a two-port system cannot be achieved without losses, as this operation would violate the second law of thermodynamics and realize a thermodynamic paradox [22],[23]. In this end-fire geometry it is the coupling to the mechanical bath that provides the required losses to block propagation in the backward direction. Indeed, for a finite pump power and $\Gamma_m \to 0$, the cooperativity approaches infinity, which, according to Eqs. (23), leads to equal intensity transmission in both directions and absence of isolation. On the other hand, if one decreases at the same rate pump power and mechanical losses, in order to keep the cooperativity constant, the non-reciprocity bandwidth reduces to zero. In the limit of zero-loss, we reach a singular condition, and again isolation disappears. Therefore, the presence of losses is necessary to achieve non-reciprocity in the transmitted intensity. This is clearly visible in Fig. 7(b), where we show that, in the absence of a pump laser, the end-fire geometry yields a pass band for light with a bandwidth given by the optical linewidth. Importantly, it is the coupling to the mechanical bath via the mechanical resonator, which comes in play when the system is pumped, that provides unidirectional losses and the resulting isolation. Although such specific end-fire geometry has the benefit of reaching optimal isolation at a relatively low cooperativity $C =$ 1, the loss mechanism in this specific situation (signaled by $\mathcal{C} = 1$) directly limits the isolation bandwidth to $2\Gamma_m$ (Fig. 7(b)), in stark contrast with the side-coupled geometry discussed in the next section.

Before concluding this section, we point the attention to a specific class of end-coupled structures, consisting of two optical waveguides resonantly coupled through a pair of identical single-mode cavities as discussed in Ref. [12],[18]. This geometry can be modeled analogously to Fig. 4a by considering the even and odd supermodes of the coupled resonators as the eigenbasis. In contrast, the localized modes of each resonator can also be considered as basis modes. Interestingly, in both cases the two modes should be driven in quadrature to achieve maximum non-reciprocal response, consistent with the general theory derived here.

## 6- Side-coupled structure

For the side-coupled structure modeled in Fig. 4(b), the scattering parameters can be calculated from Eq. (14) using the coupling matrices in (18,19). Similar to the previous case, relations (14) can be simplified when the two modes exhibit the same amount of intrinsic and external losses ($\eta_1 = \eta_2 \equiv \eta$, $\kappa_1 = \kappa_2 \equiv \kappa$), and are equally pumped, i.e., $|G_1| = |G_2| \equiv |G|$. In this case

$$S_{11} = -i\eta\kappa \frac{\mu\Sigma_m + i\hbar|G|^2 \cos(\Delta\phi)}{(\Sigma_o^2 - \mu^2)\Sigma_m - 2\hbar|G|^2 \Sigma_o}, \quad (27.a)$$

$$S_{12} = 1 - i\eta\kappa \frac{\Sigma_o\Sigma_m - \hbar|G|^2(1+\sin(\Delta\phi))}{(\Sigma_o^2 - \mu^2)\Sigma_m - 2\hbar|G|^2 \Sigma_o}, \quad (27.b)$$

$$S_{21} = 1 - i\eta\kappa \frac{\Sigma_o\Sigma_m - \hbar|G|^2(1-\sin(\Delta\phi))}{(\Sigma_o^2 - \mu^2)\Sigma_m - 2\hbar|G|^2 \Sigma_o}. \quad (27.c)$$

$$S_{22} = -i\eta\kappa \frac{\mu\Sigma_m - i\hbar|G|^2 \cos(\Delta\phi)}{(\Sigma_o^2 - \mu^2)\Sigma_m - 2\hbar|G|^2 \Sigma_o}. \quad (27.d)$$

These scattering parameters are plotted in Fig. 8 in the red-detuned regime $\bar{\Delta} = -\Omega_m$ for different pump conditions, consistent with Fig. 5. For the out-of-phase pump scenario, by increasing the pump intensity we obtain a large contrast between forward and backward transmission coefficients, at the same time increasing the isolation bandwidth of the system. It should be noted that the scattering coefficients shown in Fig. 8 exhibit similarities with those plotted in Fig. 5. In fact, a direct comparison of the expression for the scattering coefficients derived for the end-coupled and side-coupled systems (Eqs. (20,27)) shows that the two are related through the transformation

$$S_{\text{s.c.}}(\Delta\phi) = iPS_{\text{e.c.}}(\Delta\phi - \pi/2), \quad (28)$$

where in this relation $S_{\text{s.c.}}$ and $S_{\text{e.c.}}$ respectively represent the scattering matrix of the side-coupled and end-coupled structures, $\Delta\phi$ is the phase difference between pumps and $P$ is the $2 \times 2$ exchange matrix, $P = \begin{pmatrix} 0 & 1 \\ 1 & 0 \end{pmatrix}$. Equation (28) relates the transmission (reflection) coefficients of the side-coupled structure to the reflection (transmission) coefficients of the end-coupled structure when the two systems are driven with phases that differ by $\pi/2$. This relation implies that the reflection and transmission coefficients for forward and backward waves cannot be simultaneously identical, as also seen in Eqs. (20,27). Therefore, under equal-intensity pump, the left-right symmetry of this system is always broken.

### *6.1. Degenerate modes: one-way OMIT*

As in the previous example, it is of interest to explore the case of degenerate modes, i.e., $\mu = 0$. In this case, and for $\Delta\phi = \pi/2$, the transmission coefficients are simplified into

$$S_{12} = 1 - i\eta\kappa \frac{1}{\Sigma_o}, \quad (29.\text{a})$$

$$S_{21} = 1 - i\eta\kappa \frac{\Sigma_m}{\Sigma_o\Sigma_m - 2\hbar|G|^2}. \quad (29.\text{b})$$

The backward propagation is thus fully decoupled from the mechanical degree of freedom and governed only by the optical lineshape. In contrast, the forward transmission is identical to the one of a single mode optomechanical system. For forward propagation, the transmission is governed by the optical response when $G = \mathcal{G}\bar{a} \to 0$. Therefore, the system blocks light propagation over a band equal to the optical linewidth of the cavity, in both directions in the absence of a pump laser. By increasing the pump power, an optomechanically induced transparency (OMIT) signature [30]-[31] arises in the forward transmission spectrum, and for large values of $G$ the induced transparency window can be completely opened, spanning over a broad range of frequencies with peak transmission close to unity (see Figs. 9(a-c)). This operation is ultimately limited by the optical linewidth of the cavity modes [11].

### *6.2. Conditions for ideal non-reciprocity*

Equations (27) explicitly provide the conditions for ideal isolation, i.e., $S_{12} = 0$ and $S_{21} = 1$, for $\pi/2$ out-of-phase pumping. In the extreme red-detuned regime, $\bar{\Delta} = -\Omega_m$, and at optical resonance, $\omega = \Omega_m$, the transmission coefficients are

$$S_{12}(\omega = \Omega_m) = 1 - \frac{2\eta(1+2\mathcal{C})}{1+\left(\frac{\mu}{\kappa/2}\right)^2+2\mathcal{C}}, \quad (30.\text{a})$$

$$S_{21}(\omega = \Omega_m) = 1 - \frac{2\eta}{1+\left(\frac{\mu}{\kappa/2}\right)^2+2\mathcal{C}}, \quad (30.\text{b})$$

where $\mathcal{C}$ is the multi-photon cooperativity of each optical mode. Therefore, the condition to fully isolate the backward propagating probe, $S_{12}(\omega = \Omega_m) = 0$, is

$$2\left(\eta - \frac{1}{2}\right)(1 + 2\mathcal{C}) = \left(\frac{\mu}{\kappa/2}\right)^2, \quad (31)$$

which is a condition on the frequency splitting of the optical modes $\mu$ (or equivalently on the coupling rate between the two modes) in connection with the out-coupling loss ratio $\eta$ and the multiphoton cooperativity. This condition can only be satisfied for strongly coupled waveguide-cavity arrangements, i.e., $\eta > 1/2$. In addition, for degenerate modes the requirement Eq. (31) reduces to the condition of critical coupling, $\eta = 1/2$. Figure 10(a) shows the normalized mode splitting required for complete absorption of a backward propagating probe. According to Eq. (30.b), the forward transmission $S_{21}$ can become very close to unity for large cooperativities, however, it can never be equal to unity. Thus, in practice, there is always a (vanishingly small) insertion loss for the device in this side-coupled regime. The transmission contrast is shown in Fig. 10(b) in a parameter map of the normalized mode splitting and cooperativity and for a critically coupled system ($\eta = 0.5$). In this case, it is again worth exploring a scenario with no internal optical dissipation, i.e., $\eta = 1$. According to Eqs. (30), at the asymptotic limit of $\Gamma_m \to 0$, or equivalently $\mathcal{C} \to \infty$, the forward and backward transmission coefficients become equal in intensity. This shows again that the presence of losses is necessary in order to achieve optical isolation, which cannot be based on asymmetric reflections only.

This analysis points out a fundamental distinction between the operation of the two considered scenarios, end- and side-coupled geometries. In the side-coupled operation, it is the optical loss that leads to zero transmission, and mechanical loss is detrimental to achieve one-way transmission and thus the operation of the device as an isolator. On the contrary, in the end-coupled geometry mechanical loss blocks light in the unwanted propagation direction, and the optical loss should be as low-loss as possible. As a result, isolation at negligible insertion loss in the side-coupled geometry is possible only at very high cooperativities, resulting in a bandwidth ultimately limited by the optical linewidth [11]. Instead, the different loss mechanism in the end-coupled geometry leads to optimal isolation at much lower cooperativities, but at the cost of reduced bandwidths.

An interesting example of a side-coupled structure is the microring resonator system explored in Refs. [11],[16]-[17]. Such a system is typically analyzed in terms of clockwise (cw) and counterclockwise (ccw) modes. As each of the cw and ccw modes can leak only into one of the two ports, in such a description breaking the reciprocity requires driving one of the two modes while leaving the other mode unpumped [11]. Alternatively, one can consider a pair of even and odd modes as eigenbasis, falling within the general framework presented in this section [17].

## 7. Non-reciprocal amplification

In all examples discussed so far, we considered operation in the red-detuned regime, which is the most commonly considered in optomechanical systems for non-reciprocity and isolation. However, under the sideband resolved approximation, the formulation derived in the previous sections is directly applicable also to the blue-detuned regime, by simply choosing $\bar{\Delta} = \Omega_m$. Figure 11 shows the transmission coefficients associated with end- and side-coupled structures (Eqs. (20) and Eqs. (27)) when driven at the extreme blue-detuned regime, with the two modes pumped at $\Delta\phi = \pi/2$ phase difference. For an intermediate pump power range, large amplification can be achieved in this regime, either in the forward or backward direction, due to parametric gain. At resonance $\omega = -\Omega_m$ (recall that $\omega = \omega_p - \omega_L$), the transmission coefficients for the end-coupled structure become

$$S_{12}(\omega = -\Omega_m) = -2\eta \frac{\left(\frac{\mu}{\kappa/2}\right) + \mathcal{C}\sin(\Delta\phi)}{1 + \left(\frac{\mu}{\kappa/2}\right)^2 - 2\mathcal{C}}, \quad (32.\text{a})$$

$$S_{21}(\omega = -\Omega_m) = -2\eta \frac{\left(\frac{\mu}{\kappa/2}\right) - \mathcal{C}\sin(\Delta\phi)}{1 + \left(\frac{\mu}{\kappa/2}\right)^2 - 2\mathcal{C}}, \quad (32.\text{b})$$

while for the side-coupled geometry

$$S_{12}(\omega = -\Omega_m) = 1 - 2\eta \frac{1 - \mathcal{C}(1 + \sin(\Delta\phi))}{1 + \left(\frac{\mu}{\kappa/2}\right)^2 - 2\mathcal{C}}, \quad (33.\text{a})$$

$$S_{21}(\omega = -\Omega_m) = 1 - 2\eta \frac{1 - \mathcal{C}(1 - \sin(\Delta\phi))}{1 + \left(\frac{\mu}{\kappa/2}\right)^2 - 2\mathcal{C}}. \quad (33.\text{b})$$

Clearly, in both cases the transmittivities can be larger than unity, while the system is non-reciprocal. It should be noted that all the scattering parameters in Eqs. (32,33) involve a singularity at a critical power level, corresponding to $2\mathcal{C} = 1 + \left(\frac{\mu}{\kappa/2}\right)^2$. This shows the onset of instabilities when the system is excited at $\omega = -\Omega_m$. As we discuss in Section 9, such instability can occur both in the red and blue detuned regimes, but in the red-detuned regime it requires much larger power levels.

## 8- Sideband resolution

Our analysis so far has been based on the assumption of operation in the resolved sideband regime, for which the optical linewidth is much narrower than the mechanical frequency, thus filtering out the undesired sideband generated at $2\omega_L - \omega_p$ (see Fig. 12). In the following, we show that large non-reciprocity can also be achieved outside the resolved sideband regime, at the cost of a higher pump intensity. The general solution for this scenario can be derived from Eqs. (8,9), which take into account the effect of both sidebands. Using these equations and considering both terms of $\delta a_{1,2}(t)$ and $\delta a_{1,2}^*(t)$, the frequency domain equations governing the small signals can be written as:

$$i\begin{pmatrix}\Sigma_{o_1} & 0 \\ 0 & \Sigma_{o_2}\end{pmatrix}\begin{pmatrix}\delta a_1 \\ \delta a_2\end{pmatrix} - i\frac{\hbar}{\Sigma_m}\begin{pmatrix}|G_1|^2 & G_1 G_2^* \\ G_1^* G_2 & |G_2|^2\end{pmatrix}\begin{pmatrix}\delta a_1 \\ \delta a_2\end{pmatrix} - i\frac{\hbar}{\Sigma_m}\begin{pmatrix}G_1^2 & G_1 G_2 \\ G_1 G_2 & G_2^2\end{pmatrix}\begin{pmatrix}\delta a_1^*(-\omega) \\ \delta a_2^*(-\omega)\end{pmatrix} +$$

$$D^T \begin{pmatrix}\delta s_1^+ \\ \delta s_2^+\end{pmatrix} = 0, \quad (34)$$

where, $\delta a(\omega) = \mathcal{F}\{\delta a(t)\}$ and $\delta a^*(-\omega) = \mathcal{F}\{\delta a^*(t)\}$. Considering this latter relation along with its complex conjugate at negative frequencies, and using the input-output relations, we obtain

$$i\begin{pmatrix}L(\omega) & Q(\omega) \\ -Q^*(\omega) & -L^*(-\omega)\end{pmatrix}\begin{pmatrix}\delta A(\omega) \\ \delta A^*(-\omega)\end{pmatrix} + \begin{pmatrix}D & 0 \\ 0 & D^*\end{pmatrix}^T \begin{pmatrix}\delta S^+(\omega) \\ \delta S^{+*}(-\omega)\end{pmatrix} = 0, \quad (35)$$

$$\begin{pmatrix}\delta S^- \\ \delta S^{-*}(-\omega)\end{pmatrix} = \begin{pmatrix}C & 0 \\ 0 & C^*\end{pmatrix}\begin{pmatrix}\delta S^+ \\ \delta S^{+*}(-\omega)\end{pmatrix} + \begin{pmatrix}D & 0 \\ 0 & D^*\end{pmatrix}\begin{pmatrix}\delta A(\omega) \\ \delta A^*(-\omega)\end{pmatrix}, \quad (36)$$

where

$$\delta A = \begin{pmatrix}\delta a_1(\omega) \\ \delta a_2(\omega)\end{pmatrix}, \quad (37)$$

$$\delta S^\pm = \begin{pmatrix}\delta s_1^\pm(\omega) \\ \delta s_2^\pm(\omega)\end{pmatrix}, \quad (38)$$

$$L(\omega) = \begin{pmatrix}\Sigma_{o_1}(\omega) & 0 \\ 0 & \Sigma_{o_2}(\omega)\end{pmatrix} - \frac{\hbar}{\Sigma_m}\begin{pmatrix}|G_1|^2 & G_1 G_2^* \\ G_1^* G_2 & |G_2|^2\end{pmatrix}, \quad (39)$$

$$Q(\omega) = -\frac{\hbar}{\Sigma_m(\omega)}\begin{pmatrix}G_1^2 & G_1 G_2 \\ G_1 G_2 & G_2^2\end{pmatrix}. \quad (40)$$

Equations (35,36) can be solved for the modified scattering parameters as

$$\begin{pmatrix}\delta S^-(\omega) \\ \delta S^{-*}(-\omega)\end{pmatrix} = \left[\begin{pmatrix}C & 0 \\ 0 & C^*\end{pmatrix} + i\begin{pmatrix}D & 0 \\ 0 & D^*\end{pmatrix}\begin{pmatrix}L(\omega) & Q(\omega) \\ -Q^*(-\omega) & -L^*(-\omega)\end{pmatrix}^{-1}\begin{pmatrix}D & 0 \\ 0 & D^*\end{pmatrix}^T\right]\begin{pmatrix}\delta S^+(\omega) \\ \delta S^{+*}(-\omega)\end{pmatrix}.$$
(41)

Thus, the identical-frequency and frequency-converter scattering matrices, $S(\omega;\omega)$ and $S(\omega;-\omega)$, defined as

$$\begin{pmatrix}\delta s_1^-(\omega) \\ \delta s_2^-(\omega)\end{pmatrix} = S(\omega;\omega)\begin{pmatrix}\delta s_1^+(\omega) \\ \delta s_2^+(\omega)\end{pmatrix} + S(\omega;-\omega)\begin{pmatrix}\delta s_1^{+*}(-\omega) \\ \delta s_2^{+*}(-\omega)\end{pmatrix}, \quad (42)$$

become

$$S(\omega;\omega) = C + iD\left(L(\omega) - Q(\omega)(L^*(-\omega))^{-1}Q^*(-\omega)\right)^{-1} D^T, \quad (43)$$

$$S(\omega;-\omega) = iD\left(L(\omega) - Q(\omega)(L^*(-\omega))^{-1}Q^*(-\omega)\right)^{-1} Q(\omega)(L^*(-\omega))^{-1} D^{*T}. \quad (44)$$

Note that Eq. (43) should be compared with the scattering matrix obtained under a single sideband approximation (Eq. (13)), when replacing $L(\omega)$ with $L'(\omega) = L(\omega) - Q(\omega)(L^*(-\omega))^{-1}Q^*(-\omega)$. This latter term can be calculated as

$$L'(\omega) = \begin{pmatrix} \Sigma_{o_1}(\omega) - \frac{\hbar}{\Sigma_m}(1 + \alpha^*(-\omega))|G_1|^2 & -\frac{\hbar}{\Sigma_m}(1 + \alpha^*(-\omega))G_1 G_2^* \\ -\frac{\hbar}{\Sigma_m}(1 + \alpha^*(-\omega))G_1^* G_2 & \Sigma_{o_2}(\omega) - \frac{\hbar}{\Sigma_m}(1 + \alpha^*(-\omega))|G_2|^2 \end{pmatrix}, \quad (45)$$

where the frequency-dependent modification factor $\alpha$ is defined as:

$$\alpha(\omega) = \frac{\hbar\left(|G_1|^2 \Sigma_{o_2}(\omega) + |G_2|^2 \Sigma_{o_1}(\omega)\right)}{\Sigma_m(\omega)\Sigma_{o_1}(\omega)\Sigma_{o_2}(\omega) - \hbar\left(|G_1|^2 \Sigma_{o_2}(\omega) + |G_2|^2 \Sigma_{o_1}(\omega)\right)}. \quad (46)$$

Therefore, the same-frequency scattering matrix becomes

$$S(\omega;\omega) = C + iD \begin{pmatrix} \Sigma_{o_1}(\omega) - \frac{\hbar}{\Sigma_m}(1 + \alpha^*(-\omega))|G_1|^2 & -\frac{\hbar}{\Sigma_m}(1 + \alpha^*(-\omega))G_1 G_2^* \\ -\frac{\hbar}{\Sigma_m}(1 + \alpha^*(-\omega))G_1^* G_2 & \Sigma_{o_2}(\omega) - \frac{\hbar}{\Sigma_m}(1 + \alpha^*(-\omega))|G_2|^2 \end{pmatrix}^{-1} D^T. \quad (47)$$

Interestingly, the modified matrix $L'(\omega)$ exhibits the same type of asymmetry as $L(\omega)$, which in turn guarantees non-reciprocity. This property can be verified calculating the transmission coefficients obtained through the full solution of Eq. (47), and comparing it with the simplified solution Eq. (13), which neglects the effect of the other sideband. Figure 13 shows the transmission coefficients obtained based on these two approaches for three different values of sideband resolution ratio $\Omega_m/\kappa = 10$, 1 and 0.1. Here, the sideband resolution ratio is decreased by increasing the total optical losses $\kappa$, while the mechanical frequency is assumed to be constant. As seen in this figure, the solution obtained under the rotating wave approximation is close to the complete solution; only minor deviations occur at $\omega \approx -\Omega_m$. Interestingly, the non-reciprocal response is preserved in the unresolved sideband regime, even though the isolation contrast associated with the OMIT feature is significantly reduced. In fact, the reduction of the peak transparency is expected as the total losses are increased. Increasing $\kappa$ can nonetheless be beneficial, as significantly larger single-photon coupling rates $g_0 = \sqrt{\hbar/2m\Omega_m}\,\mathcal{G}$ have been reported outside the resolved sideband regime [24]. To compensate for the increased losses and maintain a strong non-reciprocal behavior, the pump power should be increased such that the multiphoton cooperativity of each mode remains constant. It should be noted that in the case of unresolved sidebands, whereas isolation at $\omega$ can be near-ideal, it would be accompanied by finite conversion to frequency $-\omega$. For applications where such frequency-converted transmission is detrimental, additional filtering could be warranted.

## 9. Linear eigenmode analysis

In this section, we rigorously explore the linear eigenmodes of the multimode optomechanical system. Such linear eigenmodes uniquely determine the overall behavior of the scattering parameters of the system at given power levels, and therefore allow discussing its temporal evolution and stability. Here, we first derive and compare the eigenvalues calculated under different approximations. Next, by exploring the evolution of the eigenvalues in the complex plane, we discuss the behavior of the reflection/transmission coefficients under different drive conditions. Then, we analyze the onset of instabilities at high pump powers.

Consider again the linearized dynamical equations (8,9) in the absence of external signal excitations. These equations can be rewritten in the matrix form

$$\frac{d}{dt}\begin{pmatrix}\delta a_1\\ \delta a_2\\ \delta a_1^*\\ \delta a_2^*\\ \delta p\\ \delta x\end{pmatrix} = i\begin{pmatrix}\bar{\Delta}_1 + i\kappa_1/2 & 0 & 0 & 0 & 0 & G_1\\ 0 & \bar{\Delta}_2 + i\kappa_2/2 & 0 & 0 & 0 & G_2\\ 0 & 0 & -\bar{\Delta}_1 + i\kappa_1/2 & 0 & 0 & -G_1^*\\ 0 & 0 & 0 & -\bar{\Delta}_2 + i\kappa_2/2 & 0 & -G_2^*\\ -i\hbar G_1^* & -i\hbar G_2^* & -i\hbar G_1 & -i\hbar G_2 & i\Gamma_m & im\Omega_m^2\\ 0 & 0 & 0 & 0 & -i/m & 0\end{pmatrix}\begin{pmatrix}\delta a_1\\ \delta a_2\\ \delta a_1^*\\ \delta a_2^*\\ \delta p\\ \delta x\end{pmatrix}, \quad (48)$$

where $\delta p = m\frac{d}{dt}\delta x$ represents the momentum of the mechanical mode. Assuming an ansatz of $(\delta a_1\ \delta a_2\ \delta a_1^*\ \delta a_2^*\ \delta p\ \delta x)^T = \boldsymbol{v}^T e^{-i\omega t}$, the eigenvalues $\omega$ are found as roots of the equation

$$\Sigma_m(\omega)\Sigma_{o_1}(\omega)\Sigma_{o_2}(\omega)\Sigma^*_{o_1}(-\omega)\Sigma^*_{o_2}(-\omega) - 2\hbar|G_1|^2\Sigma_{o_2}(\omega)\Sigma^*_{o_2}(-\omega) -$$
$$2\hbar|G_2|^2\Sigma_{o_1}(\omega)\Sigma^*_{o_1}(-\omega) = 0, \quad (49)$$

which is associated with the poles of the scattering coefficients when considering both optical sidebands. This equation can be much simplified when ignoring the coupling to conjugate optical fields centered at the opposite sideband. This can be seen from the large detuning between the diagonal elements 1 and 3 as well as 2 and 4 in the dynamical equations (48), which significantly reduces the energy transfer between the two sidebands for $|\bar{\Delta}_{1,2}| \gg \kappa_{1,2}$. In this regime, equations (48) reduce to

$$\frac{d}{dt}\begin{pmatrix}\delta a_1 \\ \delta a_2 \\ \delta p \\ \delta x\end{pmatrix} = i\begin{pmatrix}\bar{\Delta}_1 + i\kappa_1/2 & 0 & 0 & G_1 \\ 0 & \bar{\Delta}_2 + i\kappa_2/2 & 0 & G_2 \\ -i\hbar G_1^* & -i\hbar G_2^* & i\Gamma_m & im\Omega_m^2 \\ 0 & 0 & -i/m & 0\end{pmatrix}\begin{pmatrix}\delta a_1 \\ \delta a_2 \\ \delta p \\ \delta x\end{pmatrix}, \quad (50)$$

which leads to the characteristic polynomial

$$\Sigma_{o_1}(\omega)\Sigma_{o_2}(\omega)\Sigma_m(\omega) - \hbar\big(\Sigma_{o_2}(\omega)|G_1|^2 + \Sigma_{o_1}(\omega)|G_2|^2\big) = 0, \quad (51)$$

which is the denominator of the scattering coefficients in Eqs. (14). A further simplification can be made considering only one of the two mechanical sidebands. This can be done by reducing the order of the mechanical equation. For a high Q-factor mechanical mode, assuming operation around one of the two sidebands, i.e., $\omega \approx \pm\Omega_m$ for a red/blue-detuned system, the second-order operator governing the mechanical mode can be simplified as $\frac{d^2}{dt^2} + \Gamma_m \frac{d}{dt} + \Omega_m^2 = \mp i2\Omega_m\left(\frac{d}{dt} + \frac{\Gamma_m}{2} \pm i\Omega_m\right)$, and thus the mechanical equation of motion (9) reduces to

$$\frac{d}{dt}\delta x = \mp i\Omega_m \delta x - \frac{\Gamma_m}{2}\delta x \pm i\frac{\hbar}{2m\Omega_m}(G_1^*\delta a_1 + G_2^*\delta a_2), \quad (52)$$

The dynamical equations can now be written as

$$\frac{d}{dt}\begin{pmatrix}\delta a_1 \\ \delta a_2 \\ \delta x\end{pmatrix} = i\begin{pmatrix}\bar{\Delta}_1 + i\kappa/2 & 0 & G_1 \\ 0 & \bar{\Delta}_2 + i\kappa/2 & G_2 \\ \pm\hbar G_1^*/2m\Omega_m & \pm\hbar G_2^*/2m\Omega_m & \mp\Omega_m + i\Gamma_m/2\end{pmatrix}\begin{pmatrix}\delta a_1 \\ \delta a_2 \\ \delta x\end{pmatrix}, \quad (53)$$

which leads to the eigenvalue equation

$$\Sigma_{o_1}(\omega)\Sigma_{o_2}(\omega)\Sigma_m^{\pm}(\omega) \mp \frac{\hbar}{2m\Omega_m}\left(\Sigma_{o_2}(\omega)|G_1|^2 + \Sigma_{o_1}(\omega)|G_2|^2\right) = 0, \quad (54)$$

where $\Sigma_m^{\pm}(\omega) = \omega \mp \Omega_m + i\frac{\Gamma_m}{2}$ represents the positive/negative sideband inverse mechanical susceptibility. Note that, in relations (52-54), the upper/lower signs are associated with the red/blue-detuned regimes ($\bar{\Delta}_{1,2} \approx \mp\Omega_m$).

Figure 14 shows the evolution of the eigenvalues obtained from Eqs. (49,51,54) in the complex domain when the intracavity photon bias is increased from $|\bar{a}|^2 = 0$ to $|\bar{a}|^2 = 10^6$. Here, we consider both the red (a-c) and blue-detuned (d-f) regimes for a system with $|G_1| = |G_2|$, $\kappa_1 = \kappa_2$, and $\bar{\Delta}_{1,2} = \bar{\Delta} \pm \mu$, while all parameters are the same as in the examples of Figs. 5 and 8. As expected, given that the system investigated in this example is deeply within the resolved sideband regime, all the three approximations result in similar eigenvalues. It is worth noting that in all the three characteristic equations (49,51,54), the enhanced optomechanical coupling factors appear in absolute values. Therefore, and quite interestingly, based on these relations the phases of the pump beams do not have any influence on the poles of the system. This is due to our choice of using normal modes as the basis of the bare optical evolution matrix. In contrast, the drive phases play a role in the zeros of the scattering coefficients that control their frequency dispersion.

In general, the real and imaginary components of the poles are respectively associated with the resonance features and their linewidths. In fact, comparing the scattering coefficients of end- and side-coupled structures as shown in Figs. 5 and 8, for a given pump power level, similar resonance features can be distinguished irrespective of the relative phase of the drive lasers. In fact, these resonances follow the complex trend shown in Fig. 14. Considering first the red-detuned regime, given that for $|\omega - \Omega_m| < \kappa/2$ the three approximations lead to similar results, we focus on the eigenvalues obtained from the rotating wave approximation presented in Fig. 14(c). According to this figure, at low pump powers the two optical modes are separated by $2\mu$ on the real axis equally spaced on both sides of the mechanical mode which exhibits a much lower dissipation rate. By increasing the power, the mechanical mode hybridizes with the optical modes, moving towards each other along the imaginary axis. As a result, the mechanical linewidth is significantly enhanced, serving as a reservoir to absorb the backward propagating signal. As shown in Fig.

14(c), the imaginary part of the hybrid mechanical mode eigenvalue, and thus the rejection bandwidth of the device, is asymptotically limited by $\kappa/2$. In addition, the linewidths of the optical resonances are reduced, while their separation on the real frequency axis increases with increasing pump power. According to Fig. 8, while for low pump powers, the bandwidth of the forward probe is governed by the hybrid mechanical linewidth, at high powers it is determined by the separation of the hybrid optical modes on the real axis, which is ultimately limited by $2\Omega_m$.

In the blue-detuned regime (Fig. 14(d-f)), this scenario completely changes due to parametric amplification. In this case, by increasing the pump power optical and mechanical modes move in opposite directions on the imaginary axis. This results in an early appearance of an eigenvalue with positive imaginary part, corresponding to the onset of parametric amplification. In addition, as opposed to the case of red-detuning, by increasing the pump power, the hybrid optical mode eigenvalues travel toward each other. These two eigenvalues approach at a critical power level and then repel each other on the imaginary axis. Asymptotically, the imaginary part of one of the optical modes approaches $-\kappa/2$ while the other eigenvalue increases indefinitely. As a result, by increasing the power level the rejection bandwidth of the backward propagating probe approaches $\kappa/2$, while there is no bound on the bandwidth of the forward transmission. This analysis is perfectly consistent with the operation of the different geometries described in the previous section, and their dependence on the input power.

Before ending this section, it is worth noting that, similar to single-mode optomechanical systems (see for example [25][26][27]), this eigenmode analysis hints to the fact that parametric instabilities can also occur in the red-detuned regime at sufficiently large power levels. This can be shown through Eq. (49), which takes into account both sidebands. According to Fig. 14(a), by increasing the pump power, two eigenvalues from positive and negative sidebands move toward each other until merging at an exceptional point occurring at a very high power. Above this point, the two eigenvalues repel each other on the imaginary axis, leading to an unstable pole with positive imaginary part.

## 10- Biasing conditions

In this section, we explore the steady state response of the multimode cavity optomechanical system of Fig. 4 in order to find the necessary bias condition for the two optical modes in terms of input drives. The behavior of the modal bias fields is governed by Eqs. (6,7), which, when neglecting all time derivatives, is simplified to

$$i \begin{pmatrix} (\Delta_1 + \gamma_{11}|\bar{a}_1|^2 + \gamma_{12}|\bar{a}_2|^2 + i\kappa_1/2)\bar{a}_1 \\ (\Delta_2 + \gamma_{21}|\bar{a}_1|^2 + \gamma_{22}|\bar{a}_2|^2 + i\kappa_2/2)\bar{a}_2 \end{pmatrix} = -D^T \begin{pmatrix} \bar{s}_1^+ \\ \bar{s}_2^+ \end{pmatrix}, \quad (55)$$

where in these relations $\gamma_{11} = \frac{\hbar}{m\Omega_m^2}\mathcal{G}_1^2$, $\gamma_{12} = \gamma_{21} = \frac{\hbar}{m\Omega_m^2}\mathcal{G}_1\mathcal{G}_2$, and $\gamma_{22} = \frac{\hbar}{m\Omega_m^2}\mathcal{G}_2^2$. For a given driving condition, $\bar{s}_{1,2}^+$, Eqs. (55) can be solved numerically for the modal biases $\bar{a}_{1,2}$. Here, we follow the reverse approach in order to find the input pumps that allow biasing the two modes with same intensity but with a desired phase difference, i.e., $\bar{a}_2 = \bar{a}_1 \exp(i\Delta\phi) \equiv \bar{a}\exp(i\Delta\phi)$. The input fields can be obtained as

$$\begin{pmatrix} \bar{s}_1^+ \\ \bar{s}_2^+ \end{pmatrix} = -i(D^T)^{-1} \begin{pmatrix} (\Delta_1 + (\gamma_{11}+\gamma_{12})|\bar{a}|^2 + i\kappa_1/2) \\ (\Delta_2 + (\gamma_{21}+\gamma_{22})|\bar{a}|^2 + i\kappa_2/2)e^{i\Delta\phi} \end{pmatrix} \bar{a}, \quad (56)$$

To simplify the analysis, we assume $\mathcal{G}_1 = \mathcal{G}_2$ thus $\gamma_{11} = \gamma_{12} = \gamma_{21} = \gamma_{22} \equiv \gamma$. As before, we also assume $\kappa_1 = \kappa_2$, $\eta_1 = \eta_2$ and $\Delta_{1,2} = \Delta \mp \mu$. Under these conditions, we write

$$\begin{pmatrix} \bar{s}_1^+ \\ \bar{s}_2^+ \end{pmatrix} = \frac{\bar{a}}{\eta\kappa} \begin{pmatrix} d_{22}(\Delta - \mu + 2\gamma|\bar{a}|^2 + i\kappa/2) - d_{21}(\Delta + \mu + 2\gamma|\bar{a}|^2 + i\kappa/2)e^{i\Delta\phi} \\ -d_{12}(\Delta - \mu + 2\gamma|\bar{a}|^2 + i\kappa/2) + d_{11}(\Delta + \mu + 2\gamma|\bar{a}|^2 + i\kappa/2)e^{i\Delta\phi} \end{pmatrix}, \quad (57)$$

Based on this relation, and using the coupling matrices derived in Sections 4, the input fields required to achieve $\Delta\phi = \pi/2$, for the end-coupled structure are

$$\begin{pmatrix} \bar{s}_1^+ \\ \bar{s}_2^+ \end{pmatrix} = \frac{1}{\sqrt{\eta\kappa}} \bar{a} \begin{pmatrix} -i(\Delta + 2\gamma|\bar{a}|^2 + i\kappa/2) - \mu \\ (\Delta + 2\gamma|\bar{a}|^2 + i\kappa/2) + i\mu \end{pmatrix}. \quad (58)$$

while for the side-coupled geometry:

$$\begin{pmatrix} \bar{s}_1^+ \\ \bar{s}_2^+ \end{pmatrix} = \sqrt{\frac{2}{\eta\kappa}} \bar{a} \begin{pmatrix} \Delta - \delta\omega_0 + 2\gamma|\bar{a}|^2 + i\kappa/2 \\ -\mu \end{pmatrix}. \quad (59)$$

Given that $\mu$ can in principle be ignored in comparison with $\Delta$, Eqs. (58) and (59) imply that, in order to enforce a $\pi/2$ phase difference between the modal biases, the end-coupled structure should be excited from both channels with a $-\pi/2$ phase difference, while the side-coupled structure should be excited only from one port. This is a quite interesting and general result, consistent with several recent implementations of optomechanical isolators [17][18].

## 11- Time-domain simulations

While the previous results generally describe the steady-state response of a wide class of non-reciprocal systems based on optomechanical interactions, it is important to assess their temporal dynamics, governed by the nonlinear evolution equations (6,7). A rigorous numerical treatment of these equations is highly desirable, since it can justify the validity of the frequency domain scattering parameters obtained from the linearized system with or without making the rotating wave approximation. In addition, other important issues, such as the onset of optomechanical instabilities and the presence of higher-order sidebands, can be addressed with a rigorous numerical solution of the governing nonlinear dynamical equations. Such considerations can be important in properly devising pump and probe levels, in order to avoid unwanted nonlinear effects not captured by the linearized model described so far, and which can deteriorate the overall performance of the device.

By considering the mechanical momentum $p = m\, dx/dt$, we utilize a one-way propagating finite difference method to solve the set of nonlinear equations

$$\frac{d}{dt}\begin{pmatrix} a_1 \\ a_2 \\ p \\ x \end{pmatrix} = i \begin{pmatrix} \Delta_1 + \mathcal{G}_1 x + i\kappa_1/2 & 0 & 0 & 0 \\ 0 & \Delta_2 + \mathcal{G}_2 x + i\kappa_2/2 & 0 & 0 \\ -i\hbar\mathcal{G}_1 a_1^* & -i\hbar\mathcal{G}_2 a_2^* & i\Gamma_m & im\Omega_m^2 \\ 0 & 0 & -i/m & 0 \end{pmatrix} \begin{pmatrix} a_1 \\ a_2 \\ p \\ x \end{pmatrix} + \begin{pmatrix} d_{11} s_1^+ + d_{21} s_2^+ \\ d_{12} s_1^+ + d_{22} s_2^+ \\ 0 \\ 0 \end{pmatrix}, \quad (60)$$

where the output fields can be instantaneously obtained in terms of the inputs as well as the optical modal amplitudes according to Eq. (2). The response of this system to a single-sideband excitation probe can be explored by considering

$$s_1^+(t) = \bar{s}_1^+ + s_{01}^+ \exp(-i\omega t), \quad (61.a)$$

$$s_2^+(t) = \bar{s}_2^+ + s_{02}^+ \exp(-i\omega t), \quad (61.b)$$

where the small signal coefficients $s_{01}^+$ and $s_{02}^+$ are assumed to be much smaller than the biases $\bar{s}_1^+$ and $\bar{s}_2^+$ obtained from Eqs. (58,59). Here, we consider the side-coupled structure with parameters described in Fig. 5 and simulate the dynamics for a given time $t_0$ until the system reaches a steady state. The transmission coefficients are then obtained by calculating the Fourier contents of the output signal in both channels. Figure 15 shows the power spectrum of the input and output signals at both ports when driven from left (Figs. 15(a-d)) and right (Figs. 15(e-h)) directions with a probe signal at $\omega = \Omega_m$. In both cases, the transmission coefficients are in good agreement with the frequency domain analysis based on the linearized equations. In the case of backward excitation, a second harmonic at $2\Omega_m$ appears in the transmission coefficient as shown in Fig. 15(g). This is indeed due to the fact that for the side-coupled structure the pump bias at port 2 is much smaller than port 1 and in this example the backward signal power is comparable to the pump. As a result, the first order linearization of the dynamical equations is no longer strictly valid. This, however, does not significantly affect the performance of the device, as both harmonics in the transmitted signal carry less than 2% of the power, while the rest is attenuated. In principle, additional sidebands can be investigated by considering higher-order harmonics in the Taylor series expansion of the field and position variables, as done in [28] for a single-mode optomechanical system.

## 12 - Conclusions

The aim of this paper is to provide a general theoretical framework for optomechanical multi-mode systems yielding non-reciprocal responses, and derive general conditions for non-reciprocal light propagation in these systems. We have discussed different geometries that can realize optimal conditions for isolation and gyration in practical setups, and analyzed in detail end- and side-coupled geometries, which span a wide range of photonic structures. We showed that both setups can lead to near-ideal isolation but in different parameter regimes. This is related to the fact that the reservoir into which energy is lost has a drastically different nature in these cases. In principle, arbitrary photonic structures can be described in terms of the direct path scattering matrix $C$ as a linear combination of these two extreme scenarios, and can be therefore generally analyzed within the presented framework. Even though we explored optical modes with purely even and odd spatial symmetries, arbitrary mode profiles can be also considered by properly choosing the coupling matrix $D$. We derived analytical expressions for the scattering parameters for such arrangements, and the conditions for ideal isolation. The possibility of one-way amplification in the blue-detuned regime was also discussed. Our analysis shows that optomechanical isolation may be achieved even outside the sideband resolved regime, at the price of increased cooperativity levels. Finally, we have explored the pumping conditions of the system to yield the ideal driving requirements, and its behavior under nonlinear conditions in time domain.

Our results suggest that cavity optomechanics can provide a rich and powerful platform to realize reconfigurable non-reciprocal devices that can be externally controlled. In principle, optomechanical settings can be employed for more complex functionalities, such as circulation between an arbitrary number of ports as well as non-reciprocal and topologically non-trivial periodic structures [29]. In addition, our analysis suggests that, in order to exploit the full potential of optomechanical interactions, a proper design of the photonic circuitry is highly desirable. We envision the application of this theoretical framework in modeling and investigating the optical response of large optomechanical systems with multiple coupled optical and mechanical modes, in order to fully take advantage of the strong coupling between photons and phonons in a suitably tailored optomechanical material platform.

**Acknowledgments**

This work was supported by the Office of Naval Research, the Air Force Office of Scientific Research, and the Simons Foundation. This work is part of the research programme of the Netherlands Organisation for Scientific Research (NWO).

**Figures**

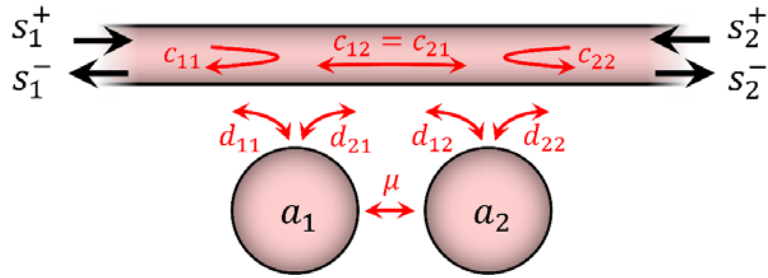

Fig. 1. Schematic representation of a two-port optical waveguide cavity arrangement involving two optical modes.

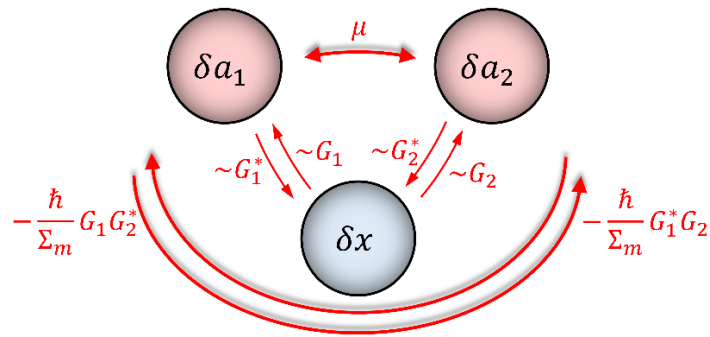

Fig. 2. The small signal model of a multimode cavity optomechanical system involving two optical modes coupled to a mechanical mode. Coupling to the mechanical resonator creates a mechanically-mediated coupling between the two optical modes, which is in general non-reciprocal.

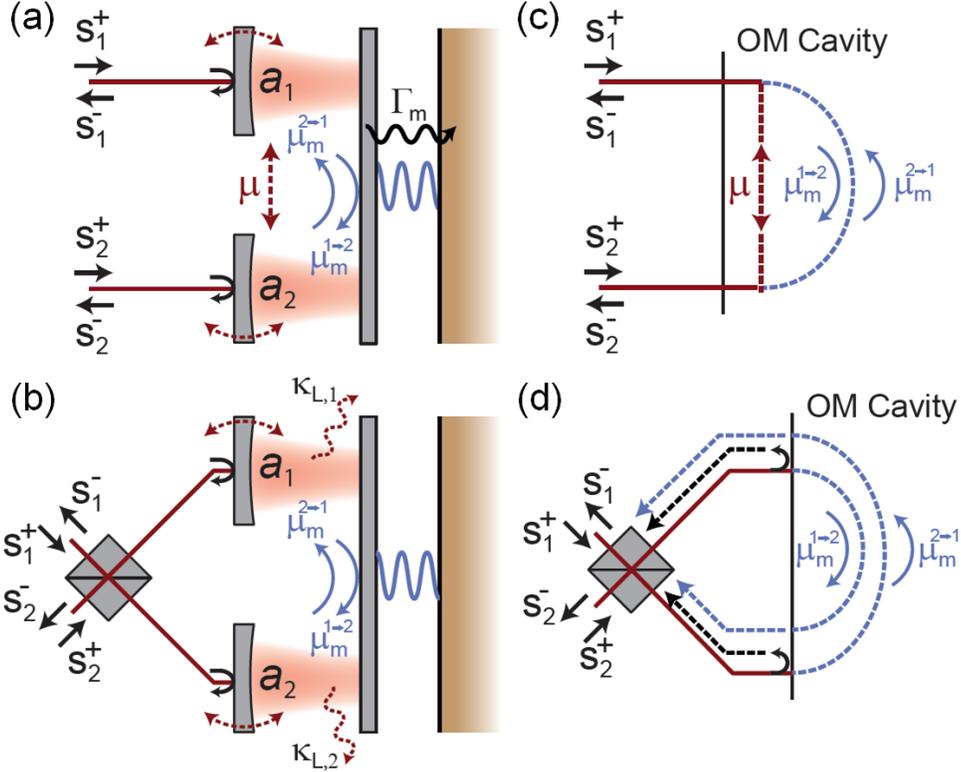

Fig. 3. Fabry-Peròt models of non-reciprocal optomechanical systems. The mechanically mediated mode conversion $\mu_m$ is in general non-reciprocal, imprinting opposite phases for photons hopping from optical mode 1 to 2, versus those hopping from mode 2 to 1. (a,c) In the absence of a direct scattering path between port 1 and 2, and the absence of direct optical coupling ($\mu = 0$), the end-fire geometry operates as a non-reciprocal phase shifter. To obtain isolation, the path that experiences a non-reciprocal phase pickup due to the mechanically mediated mode transfer needs to interfere with the direct mode coupling path ($\mu \neq 0$). Optimal isolation is achieved when the two interference paths are balanced $\mu = \mu_m$, which can fully block the signal. Note that in this scenario loss through the mechanical bath is needed to achieve isolation. This limits the operational bandwidth to the mechanical linewidth. (b,d) In contrast, in the presence of a direct-scattering path as common in a side-coupled geometry, isolation is achieved when the 'mechanically mediated path' interferes with the direct transmission between the ports. In this geometry, maximum isolation is achieved when the pump power is maximal. As in this system the non-reciprocal behaviour is fuelled by losses to the optical bath, the bandwidth is ultimately limited by the optical linewidth $\kappa$. The dashed lines in (c,d) indicate interfering optical paths.

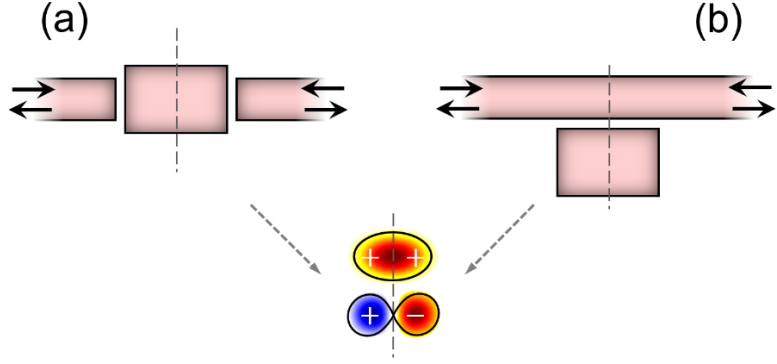

Fig. 4. Integrated photonic arrangements for (a) end-coupled (b) side-coupled systems. Inset depicts the two optical modes with even and odd symmetry.

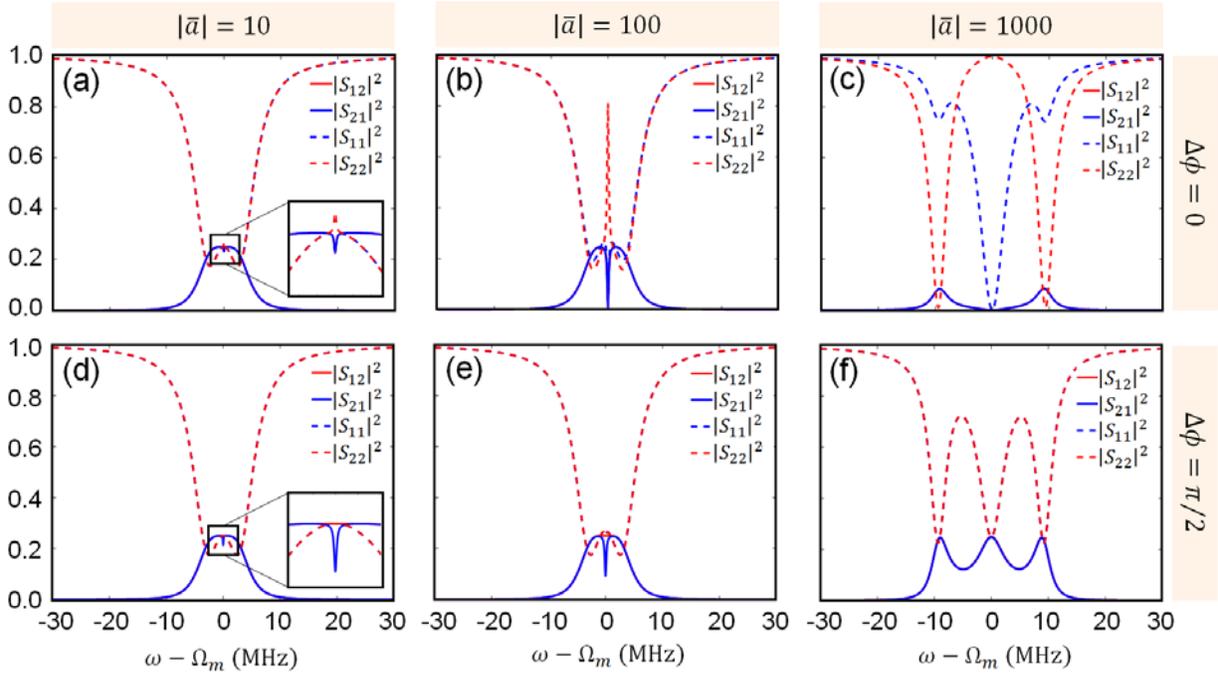

Fig. 5. Scattering parameters for an end-coupled geometry, as depicted in Fig. 4(b). The top and bottom rows are associated with $\Delta\phi = 0$ and $\Delta\phi = \pi/2$ respectively, while the intra-cavity photon number is increased from left to right. In all cases, the system is assumed to be driven in the extreme red detuned regime $\bar{\Delta} = -\Omega_m$ and the set of parameters used for this example are: $\kappa/2\pi = 1$ MHz, $\eta = 1/2$, $2\mu = 1$ MHz, $\Omega_m/2\pi = 50$ MHz, $\Gamma_m/2\pi = 10$ KHz, $m = 6$ ng, and $\mathcal{G}/2\pi = 6$ GHz/nm.

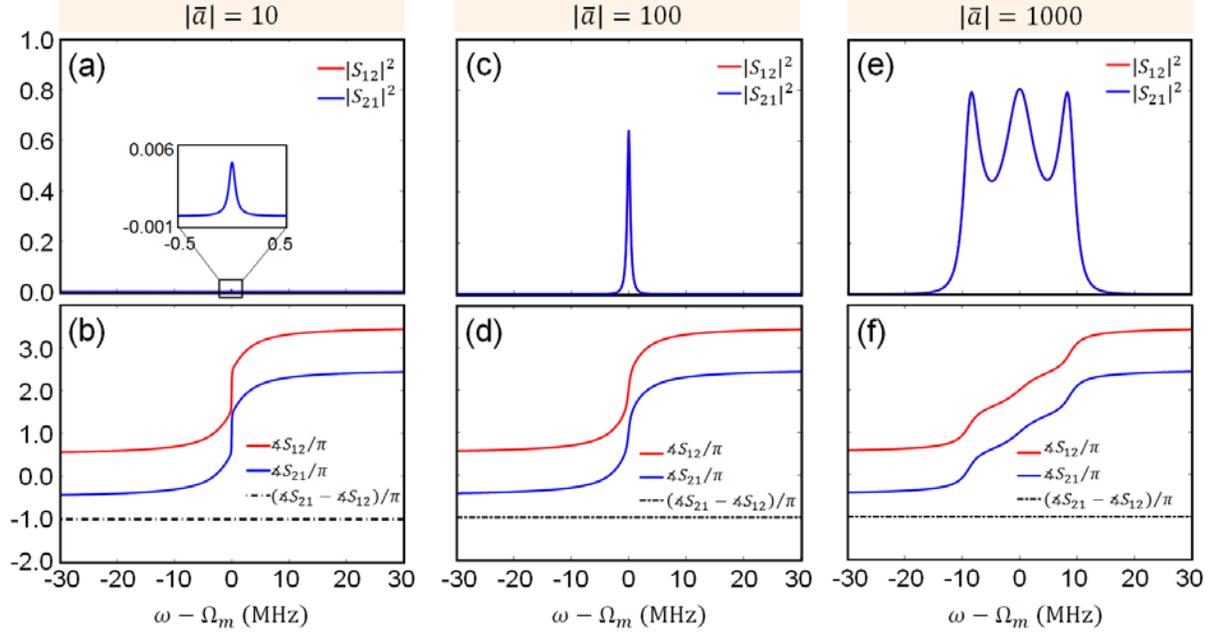

Fig. 6. Transmission coefficients of an optomechanical gyrator, obtained by removing the direct path coupling between the two optical modes, such that any coupling between the two ports is mediated through the mechanical mode. (a-f) The intensities and phases of the forward and backward transmission coefficients for different pump intensities associated with $|\bar{a}| = 10$ (a,b), $|\bar{a}| = 100$ (c,d), and $|\bar{a}| = 1000$ (e,f). In all cases, the system is assumed to be driven in the extreme red detuned regime $\bar{\Delta} = -\Omega_m$ and the set of parameters used for this example are as follows: $\kappa/2\pi = 1$ MHz, $\eta = 0.9$, $2\mu = 0$, $\Omega_m/2\pi = 50$ MHz, $\Gamma_m/2\pi = 10$ KHz, $m = 6$ ng, and $\mathcal{G}/2\pi = 6$ GHz/nm.

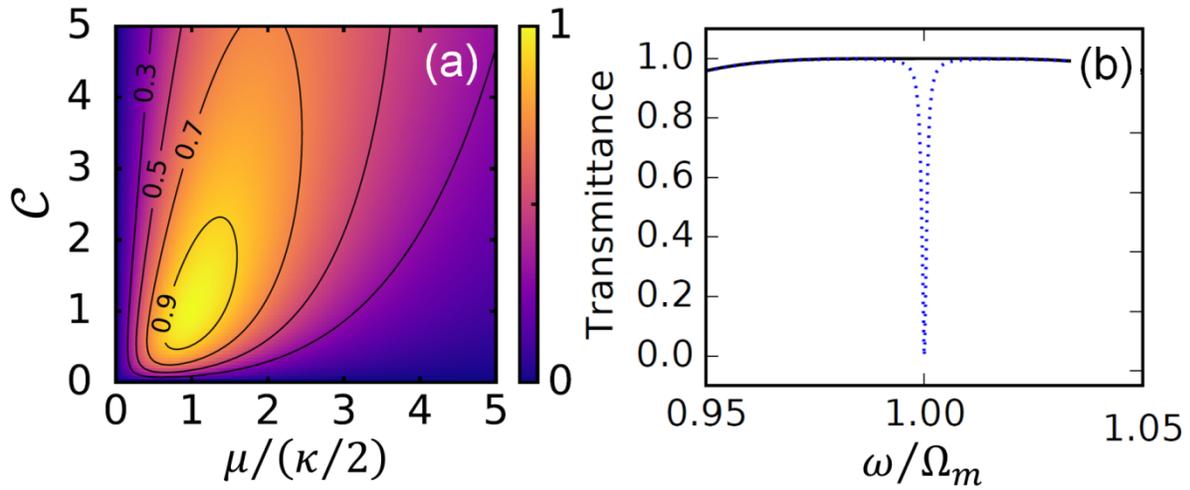

Fig. 7. (a) Maximum transmission contrast in the side-coupled structure as a function of the normalized mode splitting $\mu/(\kappa/2)$ and multi-photon cooperativity $C$. Optimal isolation is achieved for $C = \mu/(\kappa/2) = 1$ and $\eta = 1$. (b) For these optimal parameters, light in both forward (black solid line) and backward (blue dashed line) direction is transmitted over the optical bandwidth. Only in a narrow bandwidth, corresponding to the mechanical linewidth, backwards travelling light is rejected (lost in the mechanical bath), resulting in optical isolation.

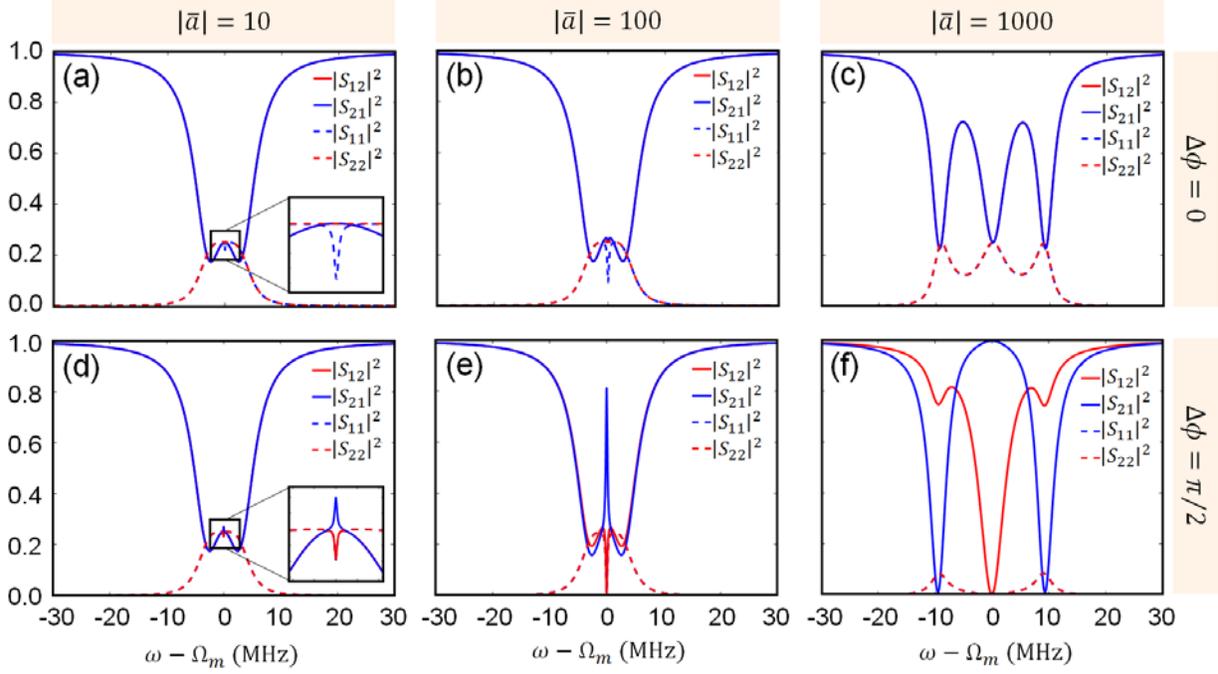

Fig. 8. Scattering parameters for a side-coupled geometry as depicted in Fig. 2(c). As in previous examples, the system is assumed to be driven in the extreme red detuned regime $\bar{\Delta} = -\Omega_m$ and the parameters used for this example are: $\kappa/2\pi = 1$ MHz, $\eta = 1/2$, $2\mu = 1$ MHz, $\Omega_m/2\pi = 50$ MHz, $\Gamma_m/2\pi = 10$ KHz, $m = 6$ ng, and $\mathcal{G}/2\pi = 6$ GHz/nm.

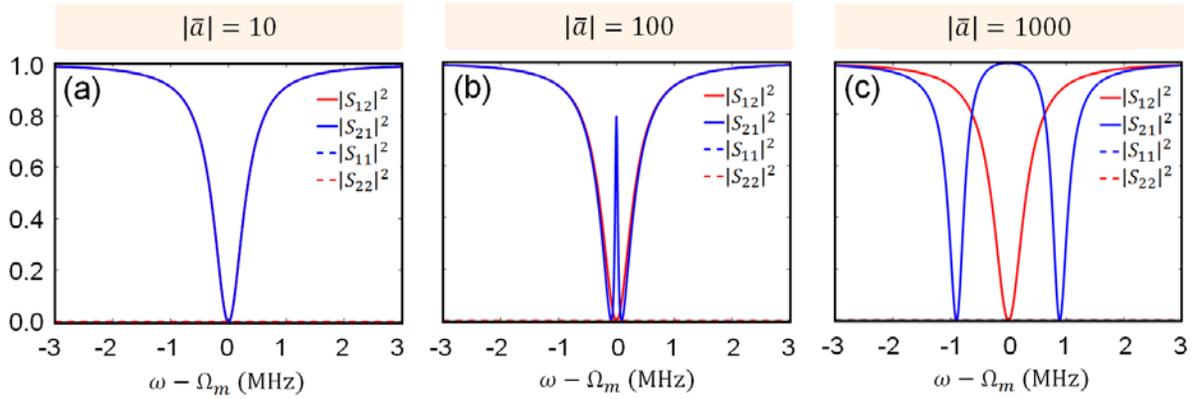

Fig. 9. Scattering parameters of the side-coupled optomechanical arrangement with degenerate optical modes for different pumping intensities. Apart from a zero mode frequency splitting $\mu = 0$, all parameters are the same as in Fig. 8.

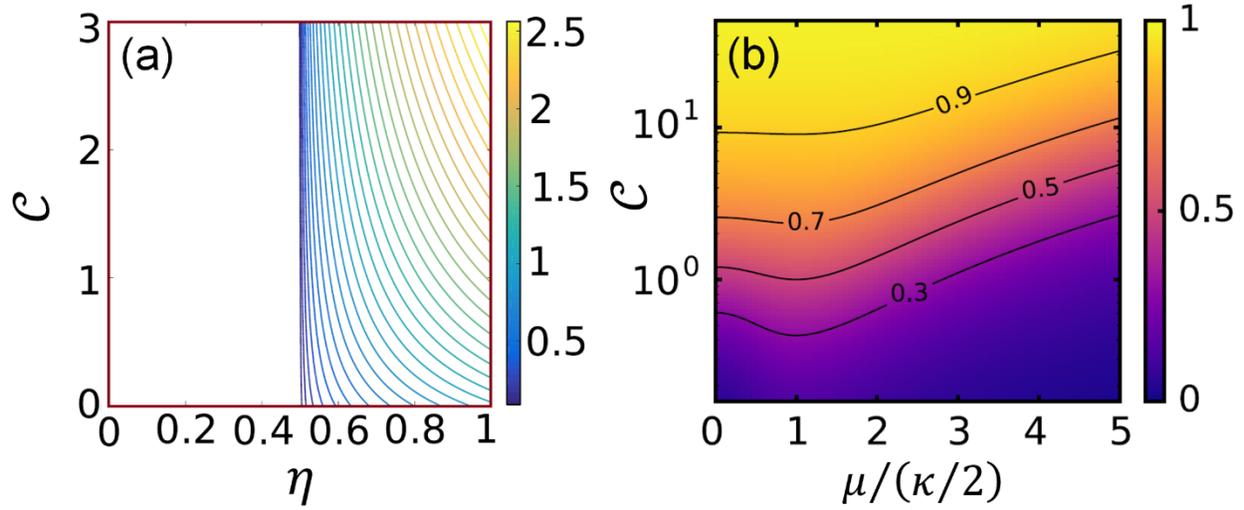

Fig. 10. (a) Normalized frequency splitting required for perfect rejection of the backward propagating probe in a side-coupled structure as a function of the outcoupling loss ratio $\eta$ and the multiphoton cooperativity of each optical mode. (b) Maximum transmission contrast as a function of the normalized frequency splitting and cooperativity for a critically coupled structure ($\eta = 0.5$).

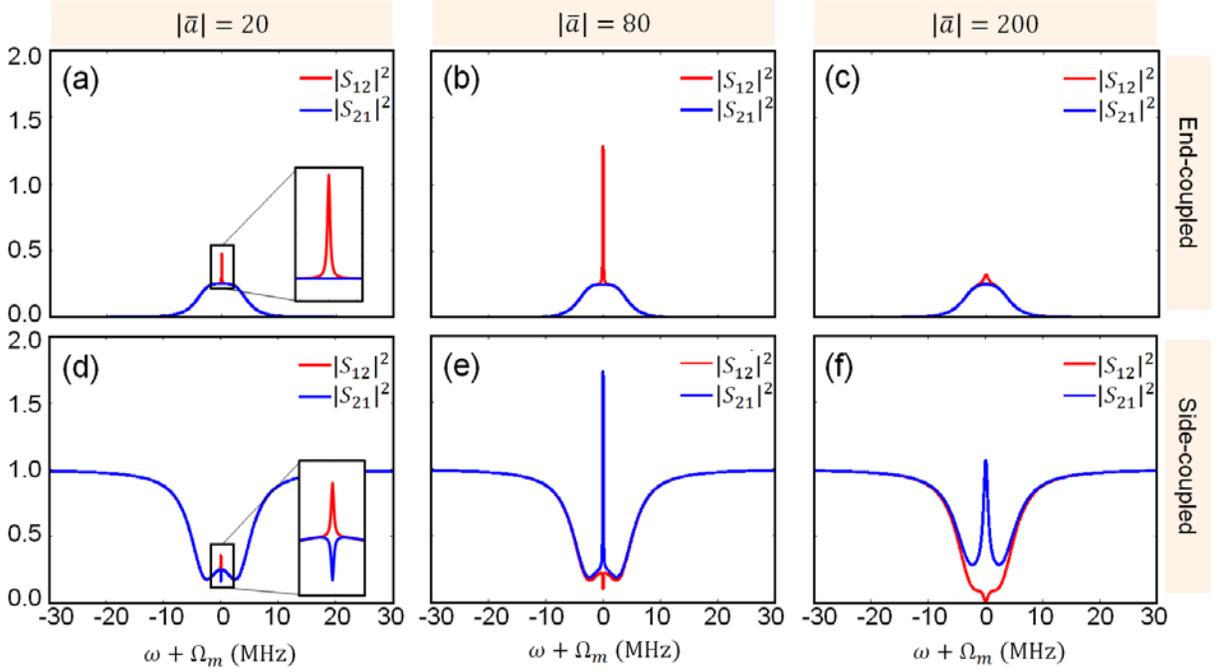

Fig. 11. Transmission coefficients of the end- (top) and side-coupled (bottom) structures when the system is driven in the extreme blue-detuned regime, i.e., $\bar{\Delta} = \Omega_m$ for different pump intensities. All parameters are the same as Figs. 5 and 8.

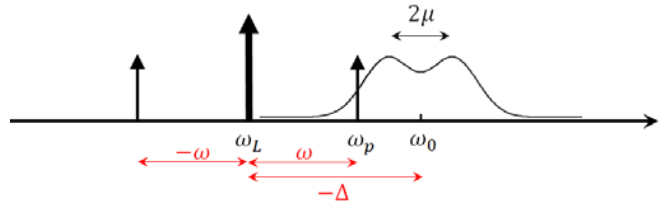

Fig. 12. A schematic illustration of the different frequency components involved in the system.

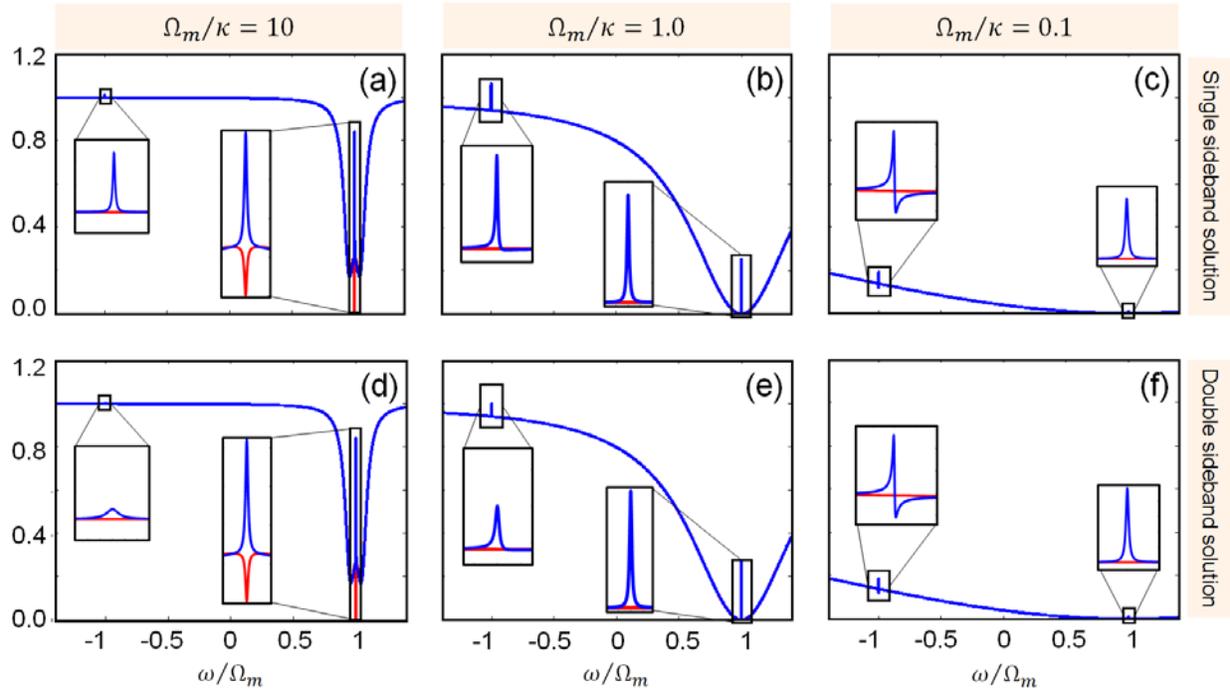

Fig. 13. The forward (blue) and backward (red) transmission coefficients for different sideband resolution ratios as obtained with (top) and without (bottom) utilizing the rotating wave approximation. Here, we have considered a side-coupled structure driven in the extreme red-detuned regime $\bar{\Delta} = -\Omega_m$ and the set of parameters used are as follows: $\eta = 0.5$, $2\mu = 5$ MHz, $\Omega_m/2\pi = 50$ MHz, $\Gamma_m/2\pi = 10$ KHz, $m = 6$ ng, $\mathcal{G}/2\pi = 6$ GHz/nm and $\bar{a} = 250$. The total optical losses are assumed to be $\kappa/2\pi = 5$ MHz (a,d), 50 MHz (b,e) and 500 MHz (c,f).

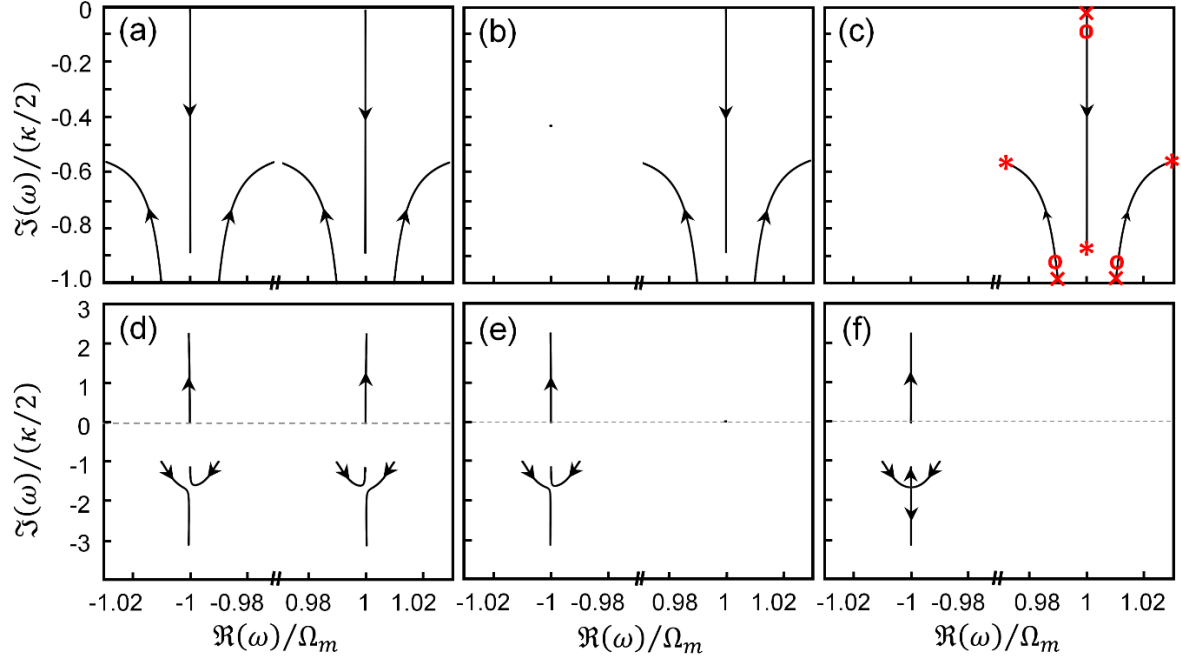

Fig. 14. Evolution of the eigenvalues of the multimode optomechanical system in the complex plane for different pump powers. (a-c) The eigenvalues obtained under the double-sideband (Eq. (49)), single-sideband (Eq. (51)), and rotating wave approximation (Eq. (54)) respectively. (d-f) The same as the top panels but for the blue-detuned regime. In all cases, the arrows show the migration direction of the eigenvalues as the pump power increases. In part (c), the markers are respectively associated with: $|\bar{a}| = 10$ (cross), $|\bar{a}| = 100$ (circle), and $|\bar{a}| = 1000$ (star). All parameters are the same as in Figs. 5 and 8.

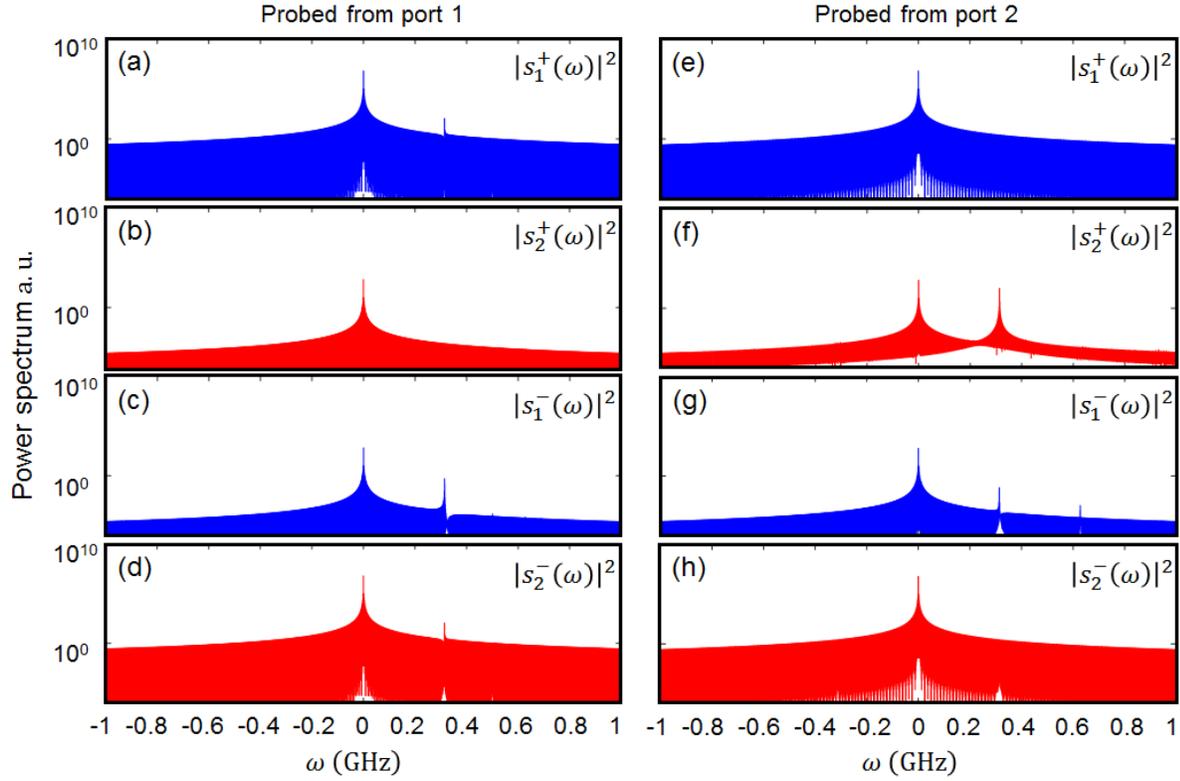

Fig. 15. Power spectrum of the input and output fields obtained from numerical solution of nonlinear dynamical equations (60) when the system is probed from the left (a-d) and right (e-h). Here we have assumed a side-coupled structure with parameters used in Fig. 5 while the probe signal is launched at $\omega = \Omega_m$ and the drive laser power is obtained from Eq. (59) such that it biases both modes with $|\bar{a}| = 1000$ and with $\pi/2$ phase difference.